\pdfoutput=1

\documentclass[11pt,twoside,a4paper,cmspaper,final,collab]{cms-tdr}
\def\svnVersion{319550}\def\cmsCernNo{2013-037}\def\cmsCernDate{\today}\def\cmsMessage{Published in the Journal of High Energy Physics as \href{http://dx.doi.org/10.1007/JHEP11(2015)071}{\doi{10.1007/JHEP11(2015)071.}}}

\begin{document}\cmsNoteHeader{HIG-14-017}
\providecommand{\hltdijmed}  {HLT\_Jet60Eta1p7\_Jet53Eta1p7\_DiBTagIP3DFastPV}
\providecommand{\hltdijhigh} {HLT\_Jet80Eta1p7\_Jet70Eta1p7\_DiBTagIP3DFastPV}
\providecommand{\hltdijvhi}  {HLT\_Jet160Eta2p4\_Jet120Eta2p4\_DiBTagIP3DFastPVLoose}

\providecommand{\hltdijlowa}{HLT\_CentralJet46\_BTagIP3D\_CentralJet38\_BTagIP3D}
\providecommand{\hltdijlow}{HLT\_CentralJet46\_CentralJet38\_DiBTagIP3D}
\providecommand{\hltdijhi}{HLT\_CentralJet60\_CentralJet53\_DiBTagIP3D}
\providecommand{\hlttrij}{HLT\_CentralJet46\_CentralJet38\_CentralJet20\_DiBTagIP3D}

\providecommand{\ldijmed}  {L1\_DoubleJetC44\_Eta1p74\_WdEta4}
\providecommand{\ldijhigh} {L1\_DoubleJetC56\_Eta1p74\_WdEta4}
\providecommand{\ldijvhi}  {L1\_SingleJet128}

\providecommand{\ldijlow}{L1\_DoubleJet36\_Central}
\providecommand{\ldijhi} {L1\_DoubleJet44\_Central}
\providecommand{\ltrij}  {L1\_TripleJet\_36\_36\_12\_Central}

\providecommand{\bb}{$b\bar{b}$}

\providecommand{\pt}{$p_T$}
\providecommand{\ptj}{$p_{T,jet}$}
\providecommand{\etaj}{$\eta_{jet}$}
\providecommand{\detajj}{$\Delta\eta_{jet,jet\prime}$}
\providecommand{\clsnf}{$95\%$ CL}

\hyphenation{had-ron-i-za-tion}
\hyphenation{cal-or-i-me-ter}
\hyphenation{de-vices}
\RCS$Revision: 302932 $
\RCS$HeadURL: svn+ssh://alverson@svn.cern.ch/reps/tdr2/papers/HIG-14-017/trunk/HIG-14-017.tex $
\RCS$Id: HIG-14-017.tex 302932 2015-09-07 09:58:19Z rmankel $

\newcommand{\CLs}{\ensuremath{\mathrm{CL_s}\xspace}}
\newcommand{\mA}{\ensuremath{m_{\mathrm{A}}}\xspace}
\newcommand{\mh}{\ensuremath{m_{\Ph}}\xspace}
\newcommand{\mH}{\ensuremath{m_{\PH}}\xspace}
\newcommand{\mPhi}{\ensuremath{m_{\phi}}\xspace}
\newcommand{\mhmax}{\ensuremath{m_{\Ph}^{\text{max}}}\xspace}
\newcommand{\mhmodp}{\ensuremath{m_{\Ph}^{\text{mod+}}}\xspace}
\newcommand{\mhmodm}{\ensuremath{m_{\Ph}^{\text{mod--}}}\xspace}

\cmsNoteHeader{HIG-14-017}
\title{Search for neutral MSSM Higgs bosons decaying into a pair of bottom quarks}

\date{\today}

\abstract{ A search for neutral Higgs bosons decaying into a
$\bbbar$ quark pair and produced in association with at least one additional b quark is
presented. This signature is sensitive to the Higgs sector of the
minimal supersymmetric standard model (MSSM) with large values
of the parameter $\tan \beta$.  The analysis is based on data
from proton-proton collisions at a center-of-mass energy of 8\TeV collected
with the CMS detector at the LHC, corresponding to an integrated luminosity of
19.7\fbinv. The
results are combined with a previous analysis based on 7\TeV data. No signal
is observed. Stringent upper
limits on the cross section times branching
fraction are derived for Higgs bosons with masses up to 900\GeV, and the results are
interpreted within different MSSM benchmark scenarios,
$\mhmax$, $\mhmodp$, $\mhmodm$,
light-stau and light-stop. Observed 95\% confidence level upper limits on $\tan \beta$, ranging from 14 to 50,
are obtained in the $\mhmodp$ benchmark scenario.}

\hypersetup{%
pdfauthor={CMS Collaboration},%
pdftitle={Search for neutral MSSM Higgs bosons decaying into a pair of bottom quarks},%
pdfsubject={CMS},%
pdfkeywords={CMS, Higgs, MSSM}}

\maketitle
\clearpage
\section{Introduction}

The discovery of a Higgs boson with a mass around
125\GeV~\cite{Aad:2012gk,Chatrchyan:2012gu,Chatrchyan:2013lba} marked
a milestone for elementary particle physics. While the measured
properties of the observed boson are in agreement with the
expectations of the standard model (SM) with the current experimental
precision, this particle could well be the first visible
member of an extended Higgs sector, which would be a direct indication
of new physics. Extended Higgs sectors are possible in various
theoretical models, such as
Supersymmetry~\cite{Wess197439,Nilles19841,Martin97,Chung:2003fi},
which relates fermionic and bosonic degrees of freedom
and in consequence requires the introduction of additional Higgs bosons
as well as a superpartner to each SM particle. The
superpartners provide potential dark-matter
candidates~\cite{Bertone2005279}, and their contribution to
quantum-loop corrections can lead to a unification of the
gauge couplings at higher energies~\cite{PhysRevD.24.1681}.
Moreover, the problem of the quadratic divergence of the Higgs boson
mass at high energies~\cite{Witten-hierarchy} is solved naturally
through cancellation of loop terms by the superpartners.

The minimal supersymmetric extension of the SM (MSSM)~\cite{Nilles19841}
contains two scalar Higgs doublets, which result in two charged Higgs
bosons, H$^{\pm}$, and three neutral ones, jointly denoted as $\phi$.
Among the latter are two CP-even (h, H) and one CP-odd state (A).  The
recently discovered boson with a mass near 125\GeV might then be
interpreted as one of the neutral CP-even states. Two parameters,
generally chosen as the mass of the pseudoscalar Higgs boson \mA and
the ratio of the vacuum expectation values of the two Higgs doublets,
$\tan \beta = v_2/v_1$, define the properties of the Higgs sector in
the MSSM at tree level.  For $\tan \beta$ values larger than one, the
couplings of the
Higgs field to down-type fermions are enhanced relative to
those to the up-type fermions. Furthermore, the A boson is nearly
degenerate in mass with either the h or H boson. These effects enhance the
combined cross section for producing these Higgs bosons in association with b
quarks by a factor of ${\approx}2\tan^2\beta$. The decay  $\phi\to\bbbar$
is expected to have a high branching
fraction ($\approx$90\%), even at large values of the Higgs boson
mass~\cite{Heinemeyer:2013tqa}.

Measurements at the CERN LHC in the
$\phi\to\tau\tau$ decay
mode~\cite{ref:Chatrchyan201268,Aad:2012cfr,Khachatryan:2014wca,Aad:2014vgg}
have lead to the most stringent constraints on $\tan\beta$ so far,
with exclusion limits in the range 4--60 in the mass interval of 90--1000\GeV.
Preceding limits had been obtained by the LEP~\cite{Schael:2006cr}
and Tevatron
experiments~\cite{Aaltonen:2009vf,Abazov:2008hu,Abazov2012569}.
Also the $\phi\to\mu\mu$ decay mode has been investigated~\cite{Aad:2012cfr,CMS:2015ooa}.
Besides extending the MSSM Higgs boson search to an independent
channel, the $\phi\to\bbbar$ decay mode is particularly
sensitive to the higgsino mass parameter $\mu$~\cite{Carena:2013qia}, and thus to
the bottom quark Yukawa coupling. In the
$\phi\to\tau\tau$ channel, the sensitivity to $\mu$ is much smaller due to a
partial cancellation of the respective radiative corrections between
the contributions to the production and
decay processes~\cite{Carena:2013qia}. Beyond the MSSM interpretation,
lepton-specific two-Higgs-doublet models (2HDM)~\cite{Branco:2011iw} may allow
for enhanced couplings of down-type quarks relative to leptons.
The $\bbbar$ decay mode is also relevant in
the more general context of exotic resonance searches, motivated
for example by dark-matter models involving mediator particles with a large
coupling to b quarks~\cite{Izaguirre:2014vva,Berlin:2014tja}.

Searches in the $\phi\to\bbbar$ decay mode have initially been
performed at LEP~\cite{Schael:2006cr} and by the CDF and D0 experiments~\cite{PhysRevD.86.091101} at
the Tevatron collider. The first and so far the only analysis at the LHC
in this channel has been performed by the CMS experiment, using
the 7\TeV data, and set significantly more stringent bounds in the mass
range  90--350\GeV~\cite{Chatrchyan:2013qga}.

In this article, the CMS search is extended by adding the data set
comprising 19.7\fbinv of proton-proton collision data, collected
at a center-of-mass energy of 8\TeV, and by the use of a refined methodology.
The higher integrated luminosity as well as the greater center-of-mass
energy allow extension of the search up to a mass of 900\GeV.

\begin{figure}[htbp]
  \centering
    \includegraphics[width=0.28\textwidth]{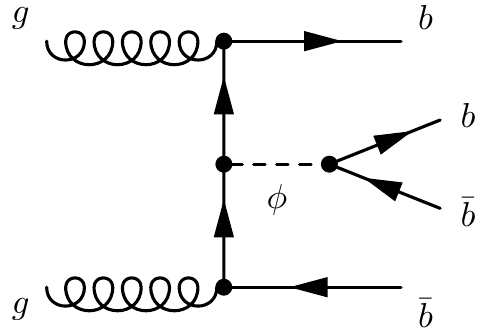}\hfill
    \includegraphics[width=0.28\textwidth]{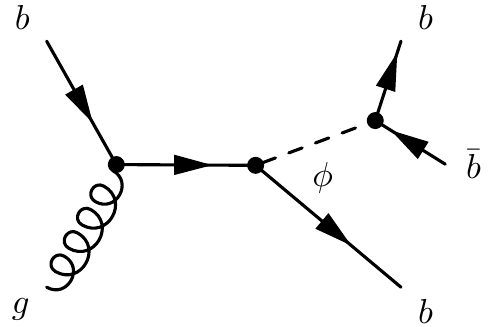}\hfill
    \includegraphics[width=0.28\textwidth]{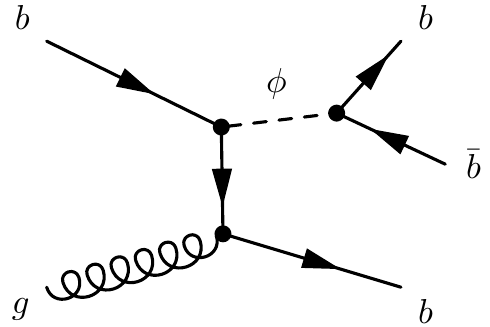}

  \caption{Example Feynman diagrams of the signal processes.
  }
   \label{fig:signalprocesses}
\end{figure}

The search is performed for neutral MSSM Higgs bosons $\phi$ with masses
$\mPhi \geq 100\GeV$ that are produced in association with at
least one b quark and decay to $\bbbar$;
an illustration of the signal process is given by the diagrams in
Fig.~\ref{fig:signalprocesses}.  The signal is thus searched for in final
states characterized by at least three b-tagged jets. No requirement
of a fourth b-tagged jet is made, since its kinematic distributions
extend significantly beyond the available acceptance, 
and the resulting signal efficiency would be very low.
Events are selected by
specialized triggers that identify b jets already at the online level.
This is important to suppress the large rate of multijet production at
the LHC. The analysis searches for a peak in the invariant mass
distribution of the two b jets with the highest \pt values, which
are assumed to originate from the Higgs boson decay.  The dominant
background is the production of heavy-flavor multijet events
containing either three b jets, or two b jets plus a third jet
originating from either a charm quark, a light-flavor quark or a gluon, which is
misidentified as a b quark jet. For the final limits, the results of the 8\TeV analysis are
combined with the previous 7\TeV analysis~\cite{Chatrchyan:2013qga}.

\section{The CMS detector}

The central feature of the CMS detector is a superconducting solenoid
of 6\unit{m} internal diameter, providing a magnetic field of
3.8\unit{T}.  Within the field volume, the inner tracker is formed by
a silicon pixel and strip tracker.  It measures charged particles
within the pseudorapidity range $\abs{\eta}< 2.5$.
The tracker
provides a transverse impact parameter resolution of approximately $15\mum$ and
a resolution on \pt  of about 1.5\% for 100\GeV
particles. Also inside the field volume are a crystal electromagnetic
calorimeter, and a brass and scintillator hadron calorimeter. Muons are
measured in gas-ionization detectors embedded in the steel flux-return
yoke, in the pseudorapidity range $\abs{\eta}< 2.4$, with detector
planes made using three technologies: drift tubes, cathode strip
chambers, and resistive-plate chambers. Matching muons to tracks
measured in the silicon tracker results in a \pt
resolution between 1\% and 5\%, for \pt values up to 1\TeV.
Forward calorimetry extends the coverage provided by the barrel
and endcap detectors up to $\abs{\eta}<5$.  A detailed description of the CMS detector,
together with a definition of the coordinate system used and the relevant
kinematic variables,
can be found in Ref.~\cite{Chatrchyan:2008zzk}.

\section{Event reconstruction and simulation}

A particle-flow
algorithm~\cite{CMS-PAS-PFT-09-001,CMS-PAS-PFT-10-001} is used to
reconstruct and identify all particles in
the event, i.e. electrons, muons, photons, charged hadrons, and
neutral hadrons, with an optimal combination of all CMS detectors
systems.

The reconstructed primary vertex with the largest $\pt^2$-sum
of its associated tracks is chosen as the vertex of the hard interaction
and used as reference for the other physics objects.

Jets are clustered from the reconstructed particle candidates using the anti-\kt
algorithm~\cite{Cacciari:2008gp} with a distance parameter of $R = 0.5$,
and each jet is required to pass
dedicated quality criteria to suppress the impact of instrumental
noise and misreconstruction.  Contributions from additional
proton-proton interactions within the same bunch crossing (pileup)
affect the jet momentum measurement.  To mitigate this effect,
charged particles associated with other vertices than the reference primary
vertex are discarded in the jet reconstruction, and residual
contributions (\eg from neutral particles)
are accounted for using a jet-area based
correction~\cite{Cacciari2008119}. Jets originating entirely from pileup
 interactions are identified and rejected based on vertex and jet-shape information~\cite{CMS-PAS-JME-13-005}.
Jet energy corrections are derived from simulation, and are confirmed
with in situ measurements of the energy balance in dijet and $\Z/\gamma$+jet
events~\cite{Chatrchyan:2011ds}.

For the offline identification of b jets, the combined secondary
vertex (CSV) algorithm \cite{Chatrchyan:2012jua} is used. This
algorithm combines information on track impact parameters and secondary
vertices within a jet in a single likelihood discriminant that provides
a good separation between b jets and jets of other flavors.
Secondary-vertex reconstruction is performed with an inclusive vertex
search amongst the tracks associated with a jet~\cite{Waltenberger:1166320}.

Simulated samples of signal and background events, also referred to as Monte Carlo (MC) samples,
were produced using
the \PYTHIA~\cite{Sjostrand:2006za} and \MADGRAPH~\cite{madgraph} event generators
and include pileup events. The response of the CMS detector
is modeled with \GEANTfour~\cite{Agostinelli:2002hh}. The MSSM Higgs signal samples,
$\Pp\Pp\to\bbbar\phi$+X with
$\phi\to\bbbar$, were produced at leading order in the 4-flavor scheme with \PYTHIA version 6.4.12. The
$\pt$ and $\eta$ distributions of the leading associated b
jet are in good agreement with the next-to-leading order (NLO)
calculations~\cite{Dittmaier:2012vm}. The multijet background from quantum chromodynamics (QCD)
processes has been produced with \PYTHIA, while for
\ttbar+jets events the \MADGRAPH event
generator was used in its version 5.1.5.11.
For all generators, fragmentation,
hadronization, and the underlying event have been modeled using \PYTHIA with tune Z2$^{\ast}$. The most recent \PYTHIA6 Z2* tune is derived
 from the Z1 tune~\cite{Field:2010bc}, which uses the CTEQ5L parton distribution functions (PDF) set, whereas
 Z2* adopts the CTEQ6L~\cite{CTEQ} PDF set.

\section{Trigger and event selection}
\label{sec:trigger}

A major challenge to this analysis is posed by the huge hadronic interaction
rate at the LHC, and it is addressed with a dedicated trigger scheme, designed
especially to suppress the QCD multijet background.
Only events with at least two jets in the
pseudorapidity range of $\abs{\eta} \leq 1.74$ are selected. The leading jet
(here and in the following the jets are ordered by decreasing \pt)
is required to have $\pt>80\GeV$, while the subleading jet must
have $\pt>70\GeV$. Furthermore, the event is only accepted if the
absolute value of the difference in pseudorapidity between any two jets fulfilling the \pt
and $\eta$ requirements is less than or equal to 1.74. The tight online requirements on the
angular variables of the jets are introduced to reduce the trigger rates while preserving the signal
significances in the probed mass range of the Higgs bosons. At the
trigger level,
b jets are identified using an algorithm that requires at least two
tracks with high 3D
impact parameter significance to be
associated with the jet. At least two jets
within the event must meet the online b tagging criteria to be accepted
by the trigger. The efficiency of the jet-\pt requirements in the trigger are derived from the
data with zero-bias triggered
events. The online b tagging efficiencies
relative to the offline b tagging
selection are obtained from simulations of QCD events generated with \PYTHIA and scaled to account for
the different b tagging efficiencies between data and simulation. The total trigger efficiency for events
satisfying the
offline selection requirements detailed below ranges from
46--62\% over the Higgs boson mass range of 100--900\GeV.

The offline selection requires events to have the two leading
jets within $\abs{\eta}\leq 1.65$ to be fully within the pseudorapidity windows
of the trigger, and
the third leading jet within $\abs{\eta}\leq 2.2$. The three leading jets
must also pass \pt thresholds of 80, 70 and 20\GeV, respectively.
In addition, the two leading jets must have a pseudorapidity
difference of $\abs{\Delta \eta_{12}}\leq 1.4$,
because the QCD multijet background increases significantly with respect
to the expected signal with increasing $\abs{\Delta \eta_{12}}$. A minimal
pairwise separation of $\Delta R>1$ between the three leading jets,
where $\Delta R=\sqrt{\smash[b]{(\Delta\eta)^2+(\Delta\phi)^2}}$ and $\Delta\eta$
and $\Delta\phi$ are the pseudorapidity and azimuthal angle
differences (in radians) between the two jets, is imposed to suppress
background from b quark pairs arising from gluon splitting.

In the following, ``triple-b-tag'' and  ``double-b-tag'' samples are
introduced, which play crucial roles in the analysis.
The triple-b-tag sample is the basis for the signal search. It is
defined by requiring all three leading jets to satisfy a tight CSV b tagging
selection requirement at
a working point characterized by a misidentification probability for
light-flavor jets (attributed to u, d, s, or g partons) of about 0.1\% at an average jet $\pt$ of 80\GeV.
The typical corresponding efficiency for b jets is about 50--60\% in the central pseudorapidity region.
The total number of events passing the trigger and offline selections
is approximately 69~k.

The double-b-tag sample plays a key role in the estimation of the
multijet background. In this selection, only two of the three leading
jets must pass the tight CSV b tagging requirement. The total number of double-b-tag
events remaining after the trigger and offline selections
is about 2.4~M. While this definition does not explicitly exclude the
triple-b-tag events, the potential signal contribution is negligible
due to the size of the QCD multijet background in the double-b-tag
sample, and a veto would lead to distortions in the background model 
described in Section~\ref{sec:bkgdmod}.

An additional flavor-sensitive quantity, the secondary vertex mass
sum of a jet, $\Sigma M_{SV,j}$, is introduced to further improve the
separation between jets of different flavor on top of the
CSV b tagging requirement. It is defined as the sum of the invariant
masses calculated from the tracks forming secondary vertices inside a
jet, and thus provides additional separation power. The extension of
the signal mass range compared to the previous 7\TeV analysis implies that
the jets can have larger \pt, with the consequence that b-tagged jets from background events
have a higher probability to contain two heavy flavor quarks instead
of at most one.
This can occur for example if a very energetic gluon splits into a
pair of b or c quarks with a narrow opening angle.
For this reason,
b and c quark pairs merged into the same jet, labeled as ``b2'' and ``c2'', respectively,
are treated separately from the
cases of unmerged b and c quarks, labeled ``b1'' and ``c1'', respectively.

The subsequent analysis will use the secondary vertex mass information
to categorize events and to build background templates.
Therefore, the secondary vertex mass sums of the three leading jets are combined
into a condensed event b tagging estimator, $X_{123}$. The
construction of this estimator is shown in Table~\ref{tab:X123}.
 Each selected jet $j$, where $j$ is the rank of the jet in order of decreasing
\pt, is assigned an index $B_j$, which can take one of the four possible
integer values from 0--3 according to its secondary vertex mass sum value, as shown in
Table~\ref{tab:X123}~(left). For jets with no reconstructed secondary
vertex, $B_j$ is also set to zero. The definition of
these index regions is motivated by the population of the
secondary vertex mass sum by the different jet flavors.
From the three indices $B_1$, $B_2$, and
$B_3$, a combined event b tagging variable $X_{123}$ is constructed as
shown in Table~\ref{tab:X123}~(right).
\begin{table}[htbp]

  \topcaption{Left:
    Definition of the index $B_j$
    according to the value of the secondary vertex mass sum of the
    jet.
    Right: Definition of the values of the combined event b tagging estimator $X_{123}$
    for all combinations of the secondary vertex mass sum indices $B_1$, $B_2$, and
    $B_3$.}
  \label{tab:X123}

  \centering
    \begin{tabular}{ccc}

      \begin{tabular}{c|c}
      \hline
        $\Sigma M_{SV,j}$ [\GeVns{}] & $B_j$  \\
      \hline
        0--1      &   0 \\
        1--2      &   1 \\
        2--3      &   2 \\
        $>$3      &   3 \\
      \hline
      \end{tabular}
& \hspace{1cm} &
      \begin{tabular}{c|ccc}
        \hline
         \multirow{2}{*}{$B_3$} & \multicolumn{3}{c}{ $B_1+B_2$ }  \\\cline{2-4}
               &  0--1    &    2--3     &   4--6     \\
        \hline
         0--1   &   0     &     1      &    2      \\
          2    &   3     &     4      &    5      \\
          3    &   6     &     7      &    8      \\
        \hline
      \end{tabular}

    \end{tabular}

\end{table}
By definition, the event b tagging variable $X_{123}$ can assume nine possible values
ranging from $0$ to $8$. The events are then categorized according to the value
of $X_{123}$, with the rationale of having sufficient statistics in
each bin. The signal is searched
for in the two-dimensional spectrum formed by the invariant mass of
the two leading jets, $M_{12}$, and the event b tagging variable
$X_{123}$.

\section{Background model}
\label{sec:bkgdmod}
The main background for this analysis originates from QCD multijet
production, with at least two energetic jets actually containing
b hadrons, and a third jet that passes the b tagging selection but possibly
as a result of a mistag. Since this type of background cannot be
accurately predicted by MC simulation, it is estimated from the data
using control samples. The chosen method is similar to the one
used in Ref.~\cite{Aaltonen:2011nh}. The background is modeled by a
combination of templates, which are constructed from the double-b-tag
sample. Only the shape of these background templates is relevant,
since the normalization will be determined by the fit to the data.

Three categories of events are distinguished in the double-b-tag
sample, which are denoted as xbb, bxb and bbx depending on whether
the jet with the highest, second-highest or third-highest \pt is exempt from
the b-tag requirement. In this notation the three jets are referred
to in order of decreasing \pt.

From these three double-b-tag categories, background templates are
constructed by weighting each untagged jet with the b tagging
probability according to its {\it assumed} flavor. In the template
nomenclature, the convention is to indicate the assumed flavor with a capital letter,
and it can be one of the five options
Q, C1, B1, C2, and B2, where Q refers to light quarks or gluons, while
C1 and C2 refer to a jet with a single charm quark and a pair of charm
quarks, respectively.
Similarly, B1 and B2 refer to jets assumed to contain a single bottom quark
and a pair of bottom quarks, respectively.
The total number of templates is therefore 15.
Each background template is a binned distribution in the
two-dimensional space spanned by $M_{12}$, the dijet mass of the two
leading jets, and the event b tagging variable $X_{123}$. For the
construction of each template, each event is weighted with the
b tagging probability corresponding to the assumed flavor of the
untagged jet. This weight accounts for the effect of the b tagging discriminant threshold.
The b tagging probability for each flavor is determined
with simulated QCD multijet events, where the flavor selection is
based on Monte Carlo truth information.
Data/MC scale factors for the b tagging efficiencies are applied where
appropriate~\cite{Chatrchyan:2012jua}. Since the b tagging efficiency has
a characteristic dependence on \pt and $\eta$ for each flavor, the
weighting results in different shapes of the $M_{12}$ distributions.
The $X_{123}$ dimension of the templates is modeled in the following way: 
In a given $M_{12}$ bin, an event can contribute to different
$X_{123}$ bins depending on the flavor of its jets and its kinematics. For the two
b-tagged jets, the secondary vertex mass sum information is taken as
measured. For the untagged jet, each of the four possible values of
the secondary vertex mass sum index is taken into account with a
weight according to the probability that a jet with given flavor, \pt
and $\eta$ will assume this value.
These probabilities,
parametrized as a function of the jet $\pt$ and $\eta$, were
determined from simulated events, and validated in control
data samples.

\begin{figure}[htpb]
  \centering
  \includegraphics[width=0.48\textwidth]{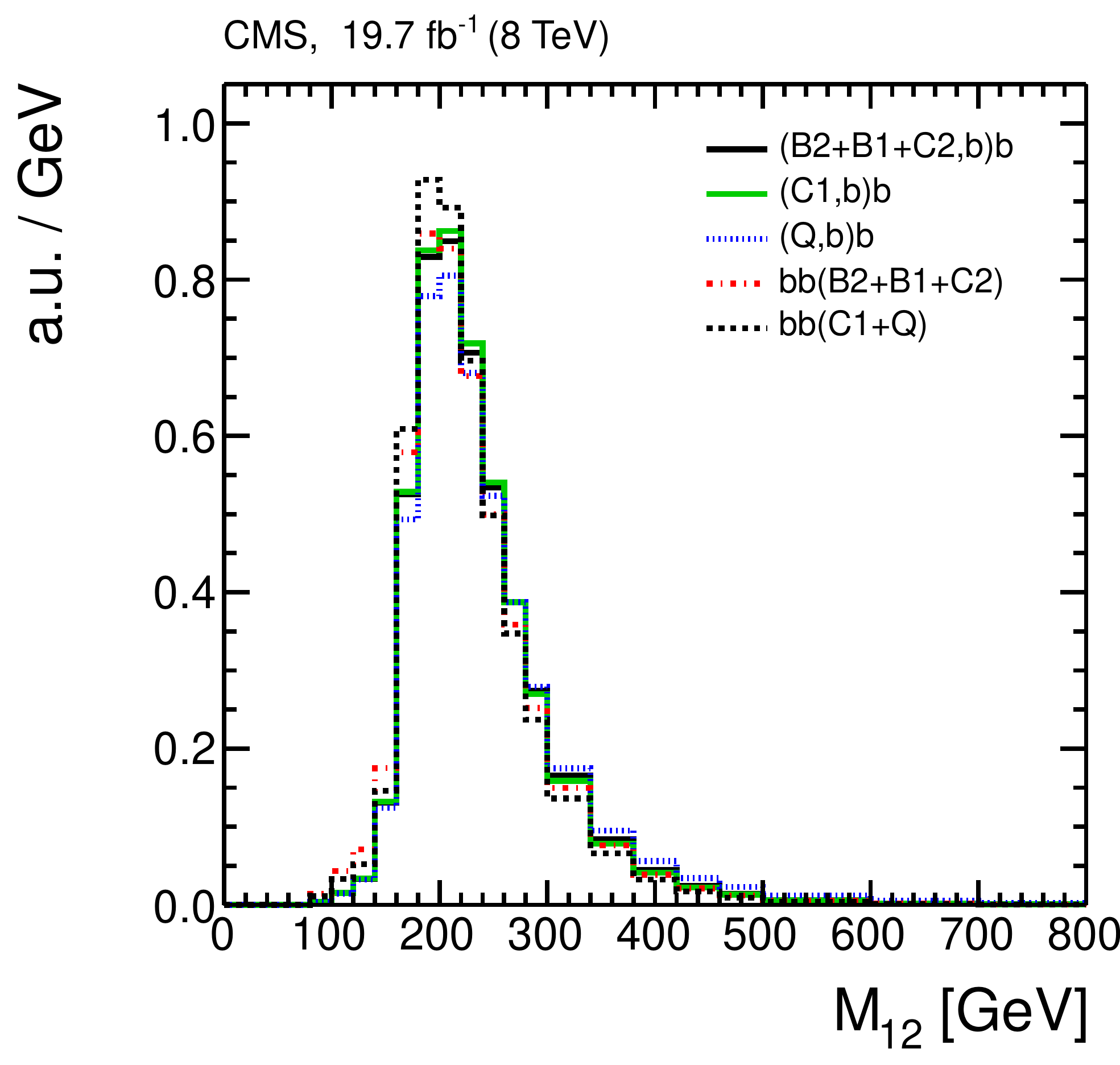}\hfill
  \includegraphics[width=0.48\textwidth]{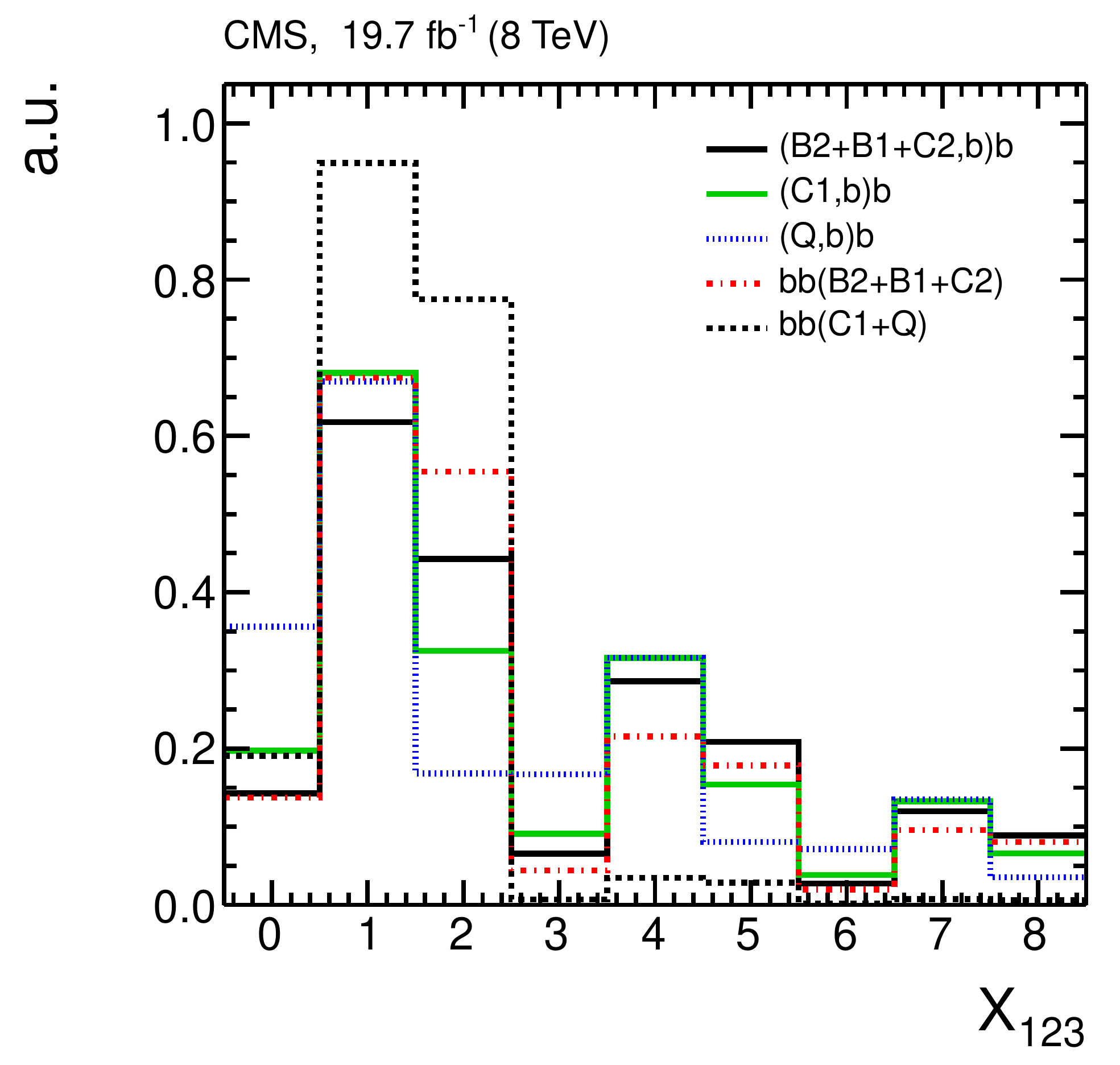}\\
  \caption{Projections of $M_{12}$ (left) and $X_{123}$ (right)
    for the five background templates used in the fit. The vertical scale is shown in arbitrary units.}
  \label{fig:templatesHighMass}
\end{figure}

Two additional corrections are applied to the templates. The first
correction addresses a contamination in the double-b-tag sample from
non-bb events at the level of a few percent. This contamination is
estimated directly from the data using a negative b tagging
discriminator~\cite{Chatrchyan:2012jua} constructed with a
track counting algorithm based on the negative impact parameter of the
tracks, ordered from the most negative impact parameter significance
upward.  A second correction is required since the online b tagging
patterns are different in the double- and the triple-b-tag samples.
In double-b-tag events, the two online b tags usually coincide with
the offline b tags, while in triple-b-tag events the online b tags can
be assigned to any two-jet subset of the three leading jets.
The correction is computed from simulated QCD multijet events, and is applied in
the form of additional weights to the events in the double-b-tag
sample.

Similarity in shape between some templates leads to unnecessary
redundancy. For this reason, similar templates are combined
using a
$\chi^2$-based metric to guide the decisions.
The relative weights in a combination are taken from MC.
In the cases where one
of the two leading jets is untagged, and the flavor assumption is the
same, e.g. Qbb, and bQb, the templates are combined, resulting in a
merged template (Q,b)b = Qbb + bQb. By analogy, also (C1,b)b, (B1,b)b,
(C2,b)b, and (B2,b)b are obtained. The resulting set of ten templates
still shows many similarities. For this reason, (B1,b)b,
(B2,b)b, and (C2,b)b are combined into a single template; bbB1, bbB2, and bbC2 into a
second; and bbC1 and bbQ into a third. The total number of templates
to be fitted in combination to the data is thus reduced
to five, namely (B2+B1+C2,b)b, (C1,b)b, (Q,b)b, bb(B2+B1+C2),
and bb(C1+Q). The projections of the $M_{12}$ and $X_{123}$ variables
are shown in Fig.~\ref{fig:templatesHighMass} for these five
background templates.

Beyond QCD multijet production, top-quark pair (\ttbar)
events pose the largest
potential background to the signal topology. The requirement of three
b-tagged jets reduces this background substantially, since only two highly
energetic b-tagged jets are expected from the decays of the top
quarks. However, one of the W bosons can decay into a $\PQc\PAQs$
pair, and the c jet can be mistagged as b jet. Using the
\ttbar Monte Carlo sample, the \ttbar contribution is found to be
relatively small; the number of \ttbar events passing the
selections of the double- and triple-b-tag datasets it estimated to be about a
factor of 70 smaller than the total
amount of data in these samples. The
invariant mass spectrum from \ttbar is very similar to the one
from the QCD multijet background, and does not show any narrow peaks.
Since the \ttbar events contribute to the double-b-tag sample, they are
also taken into account in the background model.

\section{Signal modeling}
\subsection{Signal templates}
\label{sec:signaltempl}
A signal template is obtained for each MSSM Higgs
boson mass considered by applying the full selection to the
corresponding simulated signal data set, for nominal masses in the
range of 100--900\GeV.
The sensitivity of this analysis does not extend down to cross
sections as low as that of the SM Higgs boson. Thus,
a signal model with a single mass
peak is sufficient, in contrast to the  $\phi\to\tau\tau$
analysis~\cite{Khachatryan:2014wca}, where the signal model comprises
the three neutral Higgs bosons of the MSSM, one of which is SM-like.
\begin{figure}[htbp]
  \centering
    \includegraphics[width=0.48\textwidth]{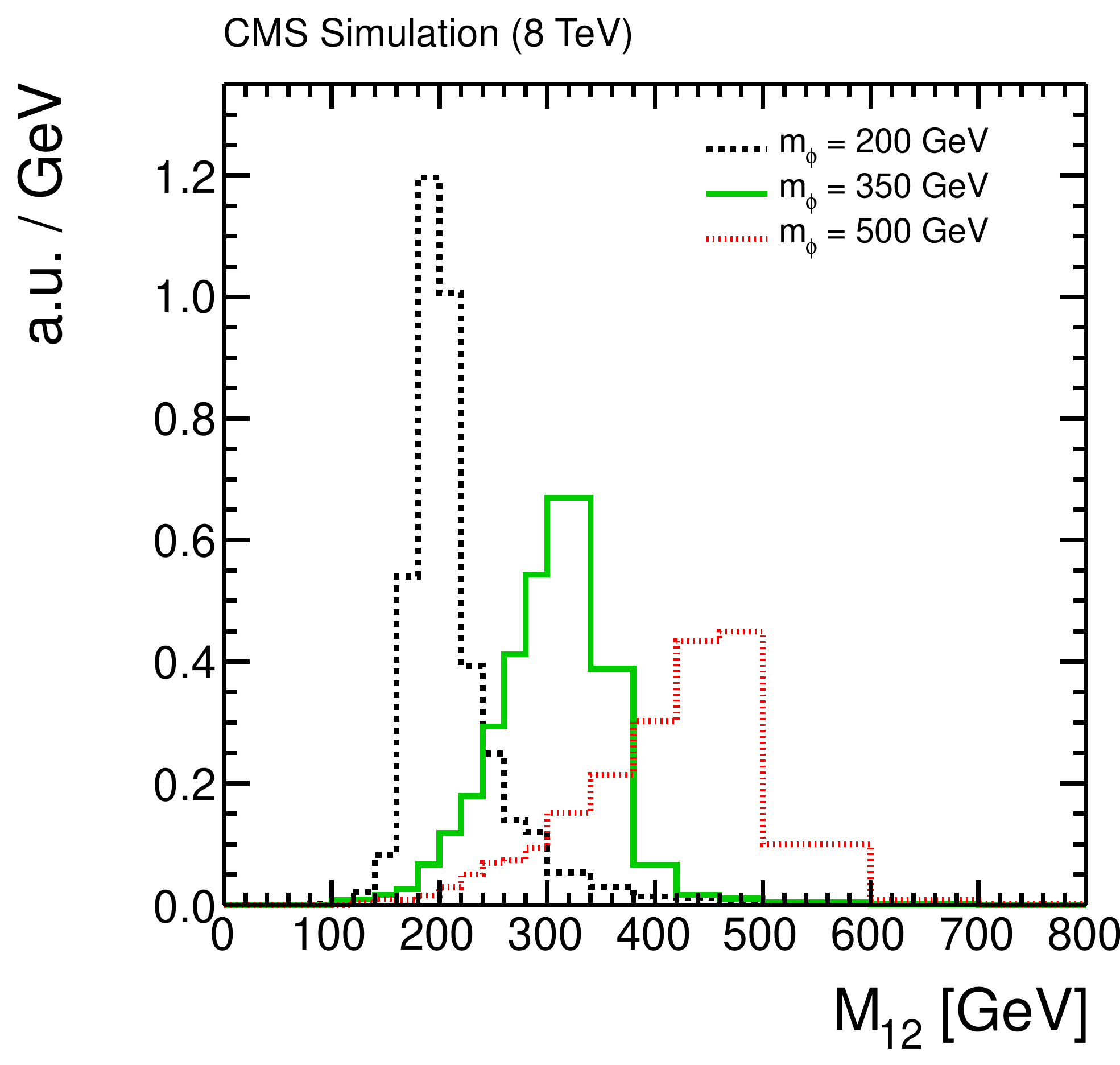}
    \includegraphics[width=0.48\textwidth]{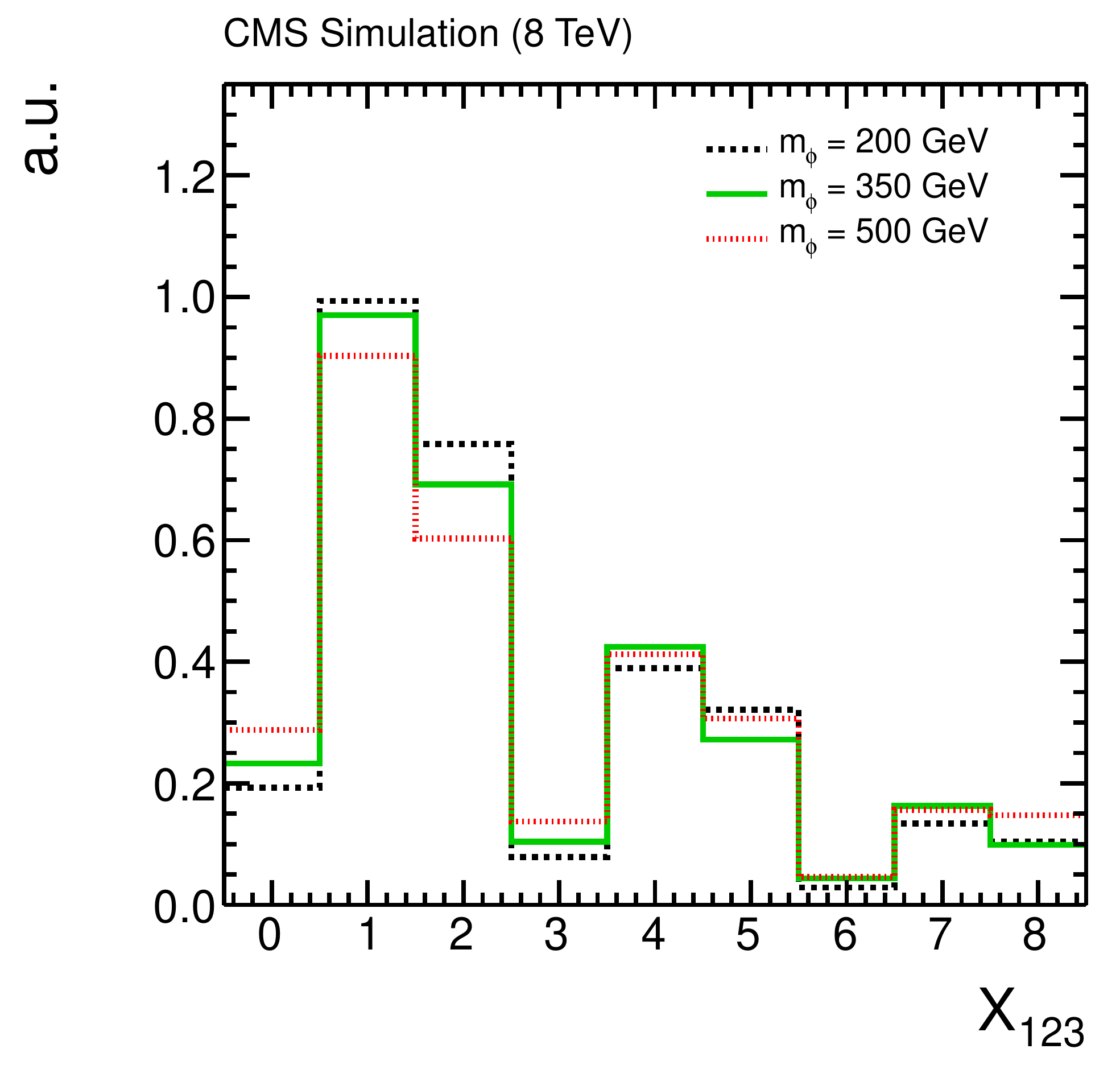}\hfill
    \caption{\label{fig:signalHTemplates} The $M_{12}$ (left) and
      $X_{123}$ (right) projections of the signal templates for Higgs
      boson masses of \mPhi= 200, 350 and 500\GeV. The vertical scale is shown in arbitrary units. }
\end{figure}
The projections for the $M_{12}$ and $X_{123}$ distributions of the
signal templates for three different Higgs boson masses are shown in
Fig.~\ref{fig:signalHTemplates}. The shape of the mass distribution
is dominated by the experimental resolution and the combinatorial
background. The natural width expected for a MSSM Higgs boson in the
considered mass and $\tan \beta$ region is negligible in
comparison with the detector resolution. At a mass of 500\GeV and
$\tan \beta=50$, for example,
the natural width of the mass peak is found to be 13\GeV, which is
only $\approx$14\% of the RMS of the reconstructed mass distribution.
The $X_{123}$ distributions show little variation with the MSSM Higgs boson mass; they
reflect the triple-b-quark signature.

\subsection{Signal efficiency}
\label{sec:signaleff}
The signal efficiency for each MSSM Higgs mass point is obtained from the simulated data sets.
The efficiency of the kinematic trigger selection has been
derived with data from control triggers and is applied by weighting. Scale factors to
account for the different b tagging efficiencies in data and MC~\cite{Chatrchyan:2012jua}
are also applied. The efficiency ranges between 0.17 and 6.38 per mille and peaks around 300\GeV.
The detailed mass dependence is shown in \ref{sec:app:efficiency}.
The decrease of the
efficiency for masses beyond 300\GeV is due to the degradation of the
b tagging efficiency at high jet \pt. For masses around 300\GeV the kinematic selections
give rise to an efficiency of approximately 0.12, which is reduced to approximately 0.0065
when triple b tagging is required.

\subsection{Fitting procedure}
\label{sec:fitting}

The overall two-dimensional distribution in the variables $M_{12}$ and $X_{123}$ is
fitted by a model combining the background templates and optionally a
signal template. A binned likelihood technique is used. The relative
contribution of each template is determined by the fit. The systematic
uncertainties are represented by nuisance parameters that are varied
in the fit according to their probability density functions.

\section{Systematic uncertainties}
\label{sec:Systematics}
The following systematic uncertainties in the expected signal and
background estimates affect the determination of the signal yield and/or its
interpretation within the MSSM.

Uncertainties in the yields of the signal contributions include the uncertainty
in the luminosity estimate~\cite{CMS-PAS-LUM-13-001}, the statistical uncertainties
in the signal MC samples, and the uncertainties of the relative online b tagging corrections.
Also taken into account are the QCD renormalization and factorization scale ($\mu_r,\ \mu_f$)
uncertainties, the uncertainties due to the parton distribution functions (PDF)
 and the strong coupling constant $\alpha_s$, and the
uncertainties in the underlying event and parton shower modeling, which all
only affect the translation of the signal cross section
into $\tan\beta$ in the MSSM interpretation. The impact
of these uncertainties on the signal acceptance is not significant.

The rate as well as the shape of the signal
contributions are also affected by the uncertainties in the trigger efficiencies, the
jet energy scale, the jet energy resolution, and
the pileup modeling, as well as the scale factors for the b-tag
efficiency, the mistag rate, and the secondary vertex mass scale.  The last
three also affect the shapes of the background
templates (recall that only the shape is relevant for the background templates).
The statistical uncertainty in the
template shape, due to the limited size of the double-b-tag sample and
due to the uncertainty in the offline b-tag efficiencies and mistag
rates, are propagated into the templates and accounted for in the
fitting procedure.  Additional systematic uncertainties in the shapes of the
background-templates arise from the impurity of the double-b-tag
sample and the online b tag correction to the templates. The
sources and types of systematic uncertainties and their impact on the expected limit are summarized in
Table~\ref{tab:systematics}.

\begin{table}[htbp]
\topcaption{Systematic uncertainties and their relative impact on the
  expected limit. The values represent an average over the
  mass range
  from 100--900\GeV, except for the template statistical and the offline b
  tagging (bc) uncertainties,
  where ranges are given.}
\label{tab:systematics}
\centering
  \begin{tabular}{l|c|c|c}
    \hline
    Source                       & Type         & Target & Impact  \\
    \hline
    Online b tagging             & Rate         & Signal & 11\% \\
    Integrated luminosity        & Rate         & Signal & 0.1\% \\

    Jet trigger                  & Rate + Shape & Signal & 0.1\% \\
    Jet energy scale             & Rate + Shape & Signal & 0.5\% \\
    Jet energy resolution        & Rate + Shape & Signal & 0.1\% \\

    Offline b tagging (bc)       & Rate + Shape & Signal + Background & 2--16\% \\
    Offline b tagging (udsg)     &        Shape & Background & 0.2\% \\
    Template stat. uncertainty   & Shape        & Background & 1--21\% \\
    Secondary vertex mass sum    & Shape        & Signal + Background & 0.9\% \\
    bb purity correction         & Shape & Background & 3.4\% \\
    Online b tagging correction  & Shape & Background & 0.5\% \\
    \hline
  \end{tabular}

\end{table}

\section{Results}
\label{sec:results}
\subsection{Background-only fit}

In the first step, an unconstrained fit is performed without inclusion
of a signal template, involving a linear combination of the background
templates only.
\begin{figure}[htbp]
  \centering
  \includegraphics[width=0.48\textwidth]{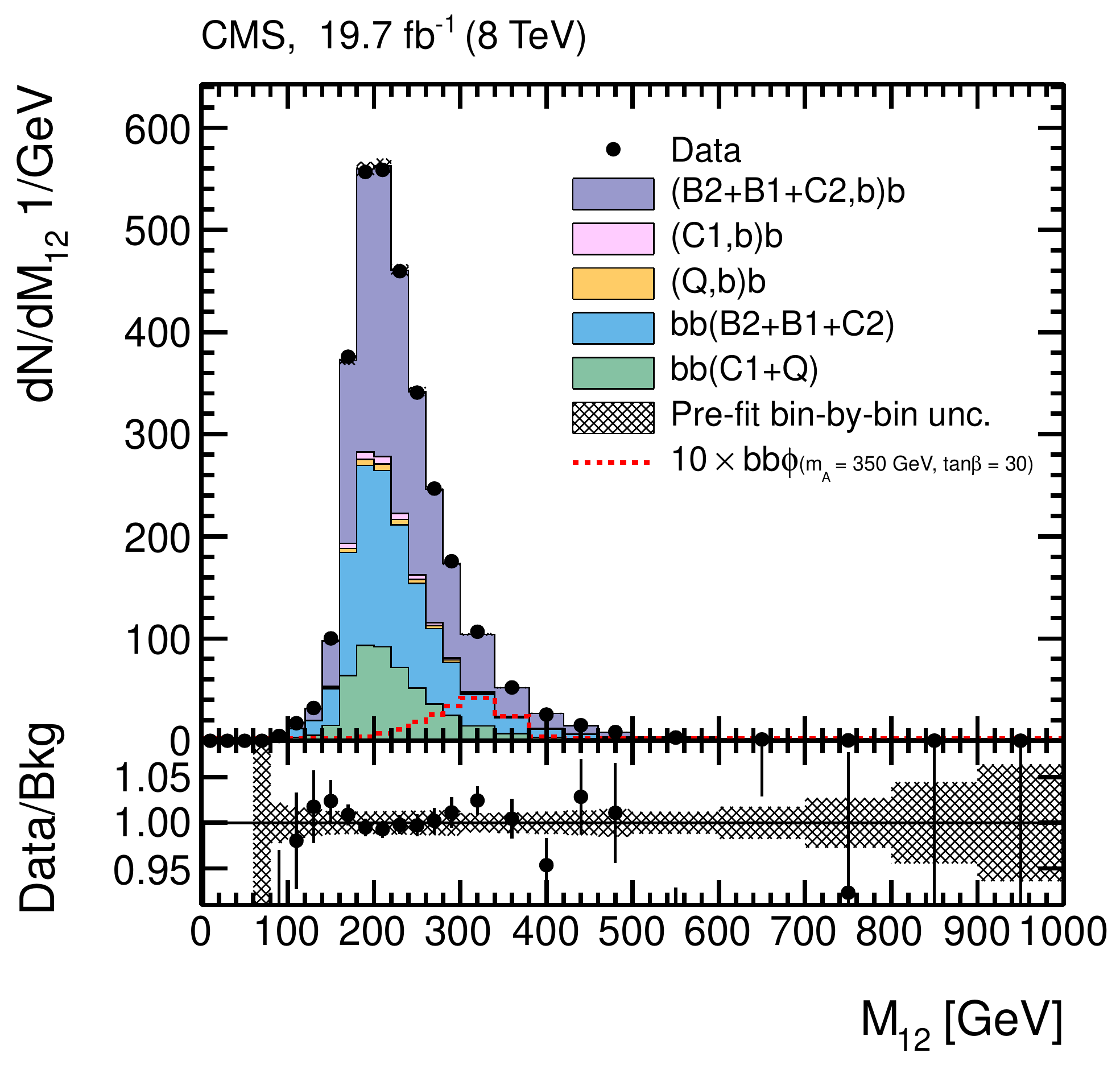}\hfill
  \includegraphics[width=0.48\textwidth]{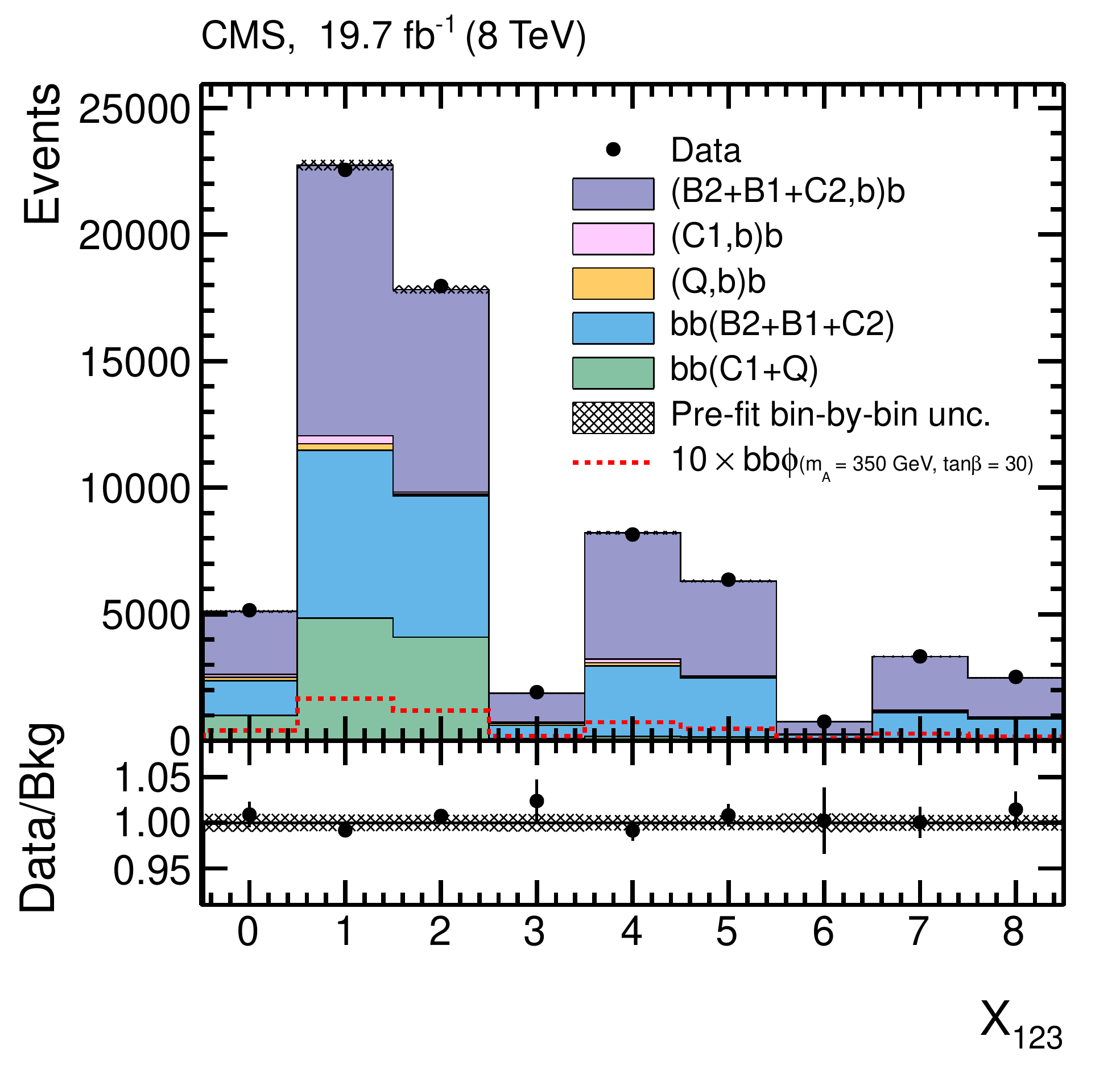}\hfill
  \caption{\label{fig:template_fit_BG} Projections of the
    dijet mass $M_{12}$ (left) and event b-tag variable $X_{123}$
    (right) in the triple-b-tag sample, together with the corresponding
    projections of the fitted background templates. The hatched area
    shows the total bin-by-bin background uncertainty of the templates
    prior to the fit, which takes into account the limited size of the
    double-b-tag sample and the uncertainties of the offline b-tag efficiencies
    and mistag rates. For illustration, the signal contribution expected in
    the \mhmax benchmark scenario of the MSSM with $\mA=350\GeV$, $\tan\beta=30$, and $\mu = +200$\GeV is overlayed, scaled by a factor 10 for better readability.
    In addition, the ratio of data to the
    background estimate is shown at the bottom.}
\end{figure}
Results are shown in Fig.~\ref{fig:template_fit_BG} and
Table~\ref{tab:fitFractions}. The template-based background
model describes the data well within the uncertainty of the template
fits with a goodness-of-fit of $\chi^2 / N_{{\rm dof}} = 207.9/209$, where $N_{\rm dof}$ is the number
of degrees of freedom, corresponding to a $p$-value of
0.51. As expected, the fit is dominated by templates involving triple
b-jet signatures, whose fitted total contributions amount to
$\approx$82\%.

\subsection{Combined fit of signal and background templates}
\begin{figure}[htbp]
  \centering
  \includegraphics[width=0.48\textwidth]{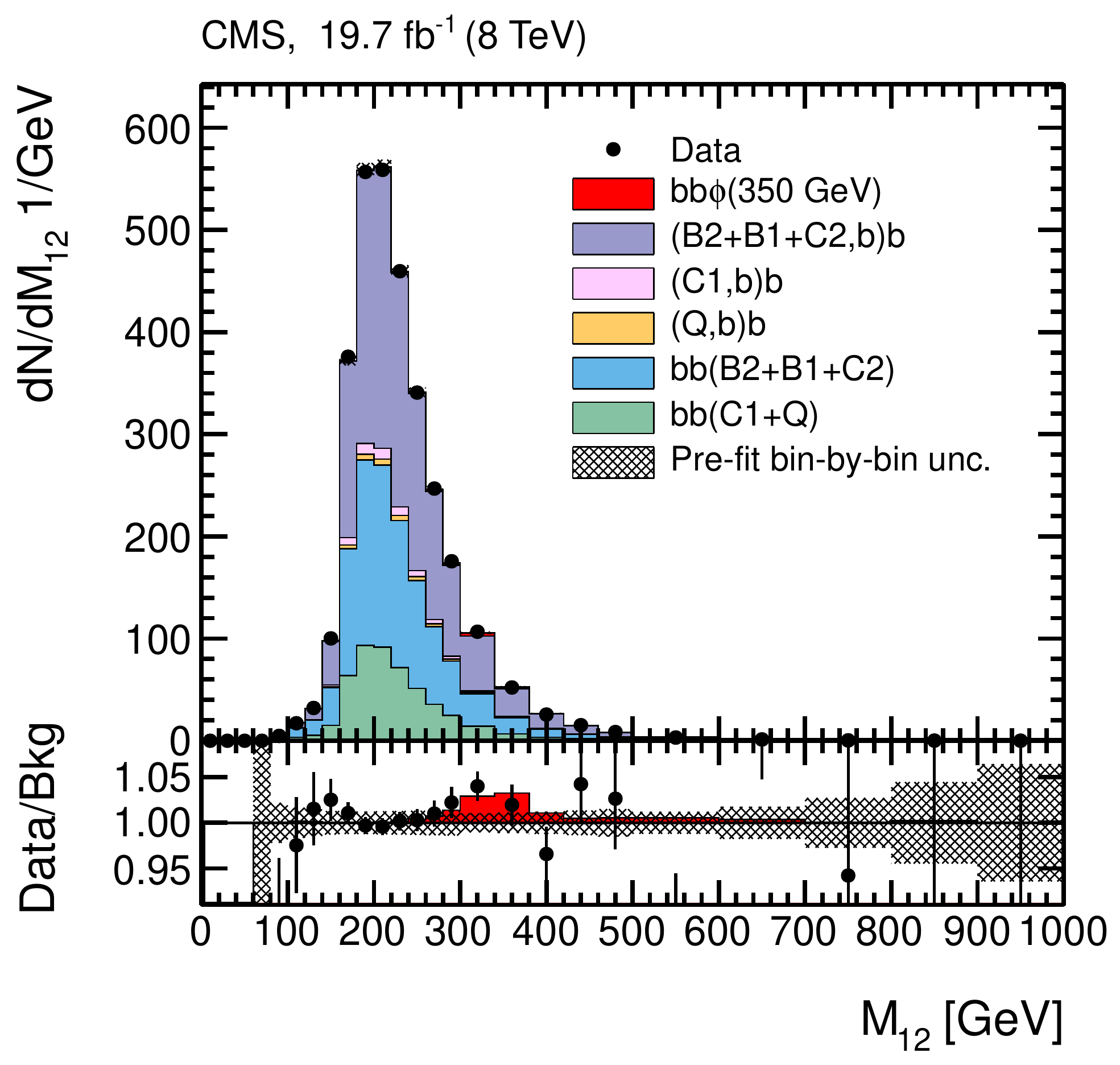}\hfill
  \includegraphics[width=0.48\textwidth]{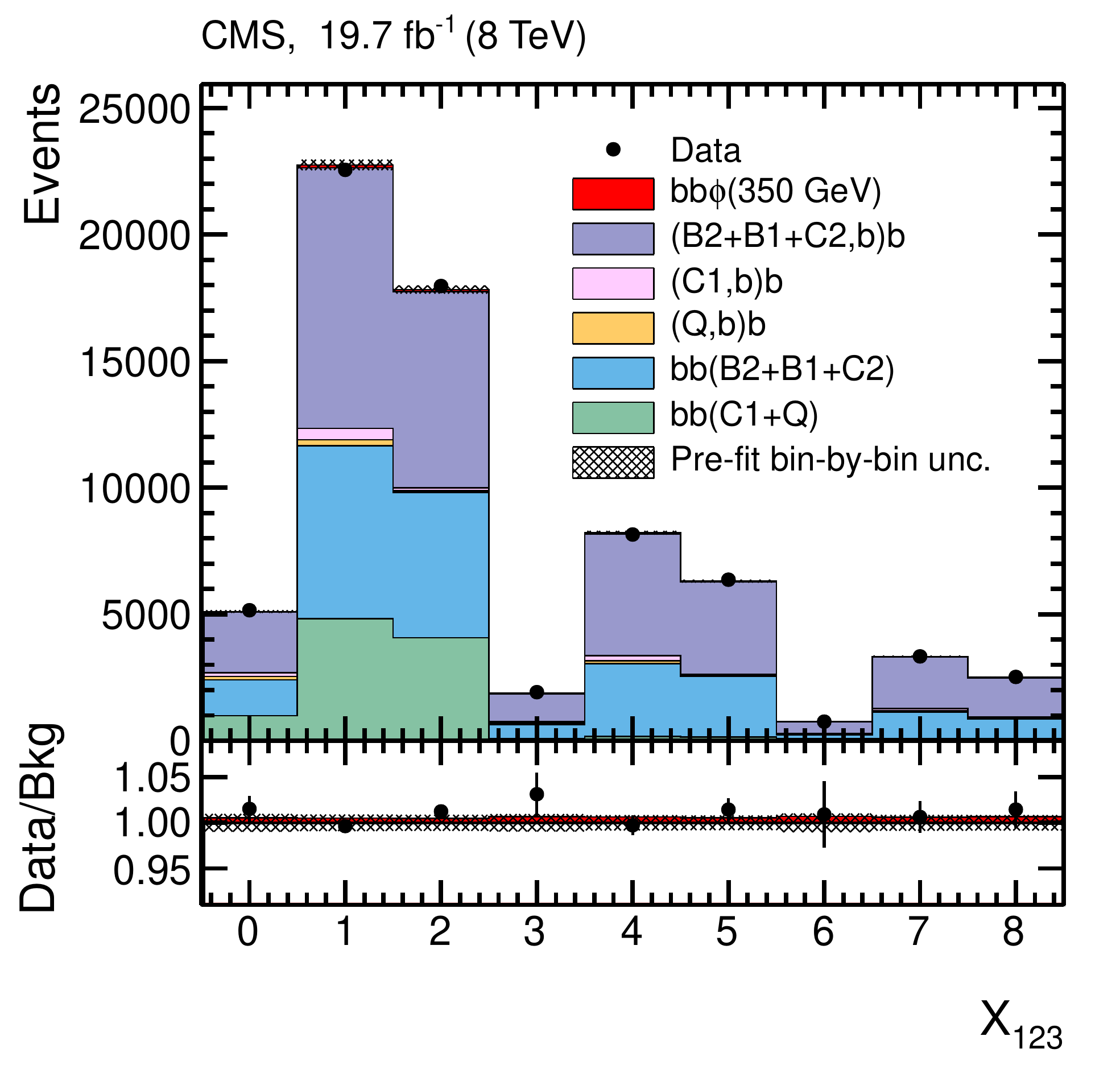}\hfill
  \caption{\label{fig:HighMass2012_m350_projections} Results from the
    combined fit of signal and background templates in the
    triple-b-tag sample, at the 350\GeV mass point. The left plot
    shows the projections of the dijet mass $M_{12}$, the right plot
    the projections of the event b-tag variable $X_{123}$.
    The red graph represents the fitted Higgs signal contribution.
    The hatched area
    shows the total bin-by-bin background uncertainty of the templates prior to the fit,
    which takes into account the limited size of the
    double-b-tag sample and the uncertainties of the offline b-tag efficiencies
    and mistag rates. In addition, the ratio of data to the
    background estimate is shown at the bottom.}
\end{figure}
In the second step, a signal template is included together with the
background templates in the fit, with the relative fractions of signal and background templates allowed to vary freely.
The fit is performed for all considered Higgs boson masses from $100$
to $900$\GeV. None of the fits shows any significant signal excess.
Results for a Higgs boson mass of $350$\GeV  are shown in
Fig.~\ref{fig:HighMass2012_m350_projections} and Table~\ref{tab:fitFractions}. At this mass point, the highest
fluctuation in the fitted Higgs boson production cross section is observed,
corresponding to a local significance of approximately 1.5 standard
deviations. The goodness-of-fit is $\chi^2 / N_{\mathrm{dof}} =
205.2/208$, corresponding to a $p$-value of 0.54.
\begin{table}[htbp]
  \topcaption{Relative contributions of the individual templates as
    determined by the background-only and by the signal+background fit
    for a Higgs boson mass hypothesis of 350\GeV.}
  \label{tab:fitFractions}
  \centering
    \begin{tabular}{c|r@{$\,\pm\,$}l|r@{$\,\pm\,$}l}
      \hline
       \multirow{2}{*}{Template}        & \multicolumn{2}{c|}{Background-only fit} & \multicolumn{2}{c}{Signal+background fit} \\
      & \multicolumn{2}{c|}{fraction [\%]} & \multicolumn{2}{c}{fraction [\%]} \\
      \hline
      (B2+B1+C2,b)b & \ \ \ \ \ \ \ \ \  51.3 & 3.5 & \,\ \ \ \ \ \ \ \ \ \ \  49.5 & 3.9 \\
      (C1,b)b       &  1.3 & 2.3 &  1.7 & 3.1 \\
      (Q,b)b        &  1.2 & 2.0 &  1.1 & 1.5 \\
      bb(B2+B1+C2)  & 31.2 & 3.2 & 32.2 & 3.4 \\
      bb (C1+Q)     & 15.1 & 0.9 & 15.0 & 0.9 \\
      \hline
      bb$\phi(m=350\GeV)$ & \multicolumn{2}{l|}{\,\ \ \ \ \ \ \ \ \ \ \ \ \ \ \ \ ---} & \,\ \ \ \ \ \ \ \ \ \ \ \ 0.5 & 0.3 \\
     \hline
    \end{tabular}

\end{table}

\subsection{Upper limits on cross sections times branching fractions}
Cross sections are obtained from the fractions determined by the fit
multiplied by the total number of data events after the selection
in the signal region, and divided by the corresponding signal efficiencies
(Section~\ref{sec:signaleff}) and the integrated luminosity of
19.7\fbinv.

\begin{figure}[htbp]
  \centering
    \includegraphics[width=0.48\textwidth]{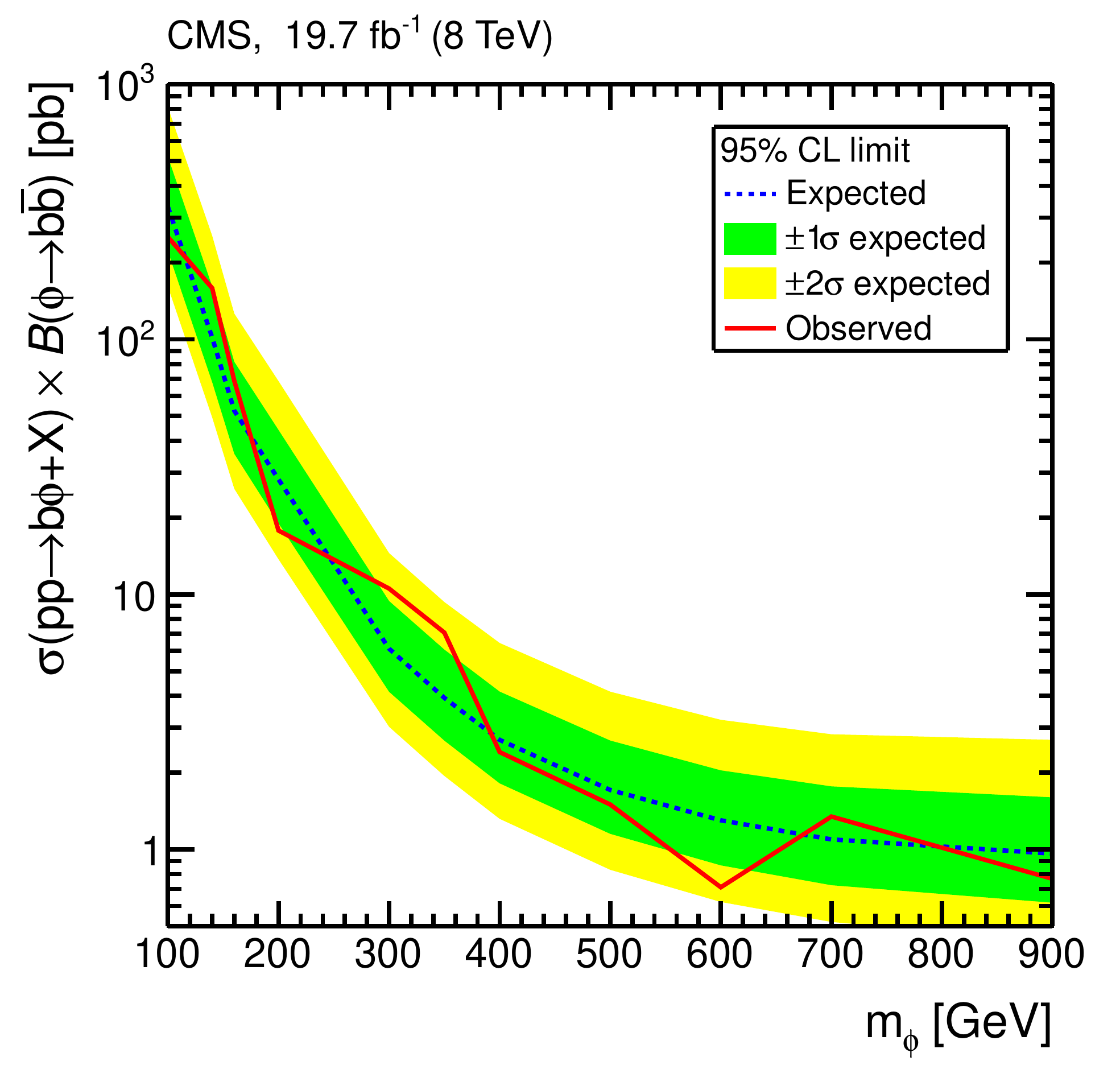}
    \caption{\label{fig:xsec_mA_bands_csvt} Expected and observed
      upper limits at \clsnf\ on
      $\sigma(\Pp\Pp\to\PQb\phi+\mathrm{X})\,\mathcal{B}(\phi\to\bbbar)$ as a function of \mPhi, where $\phi$ denotes a generic neutral Higgs-like state.}
\end{figure}
In the absence of any significant signal,
the results are translated into upper limits on the cross section
times the branching fraction, $\sigma(\Pp\Pp \to \PQb\phi + \mathrm{X}) \, \mathcal{B}(\phi \to \bbbar)$, of a generic Higgs-like state in the mass range 100--900\GeV. For
calculations of exclusion limits, the modified frequentist
construction \CLs \cite{Junk:1999kv,Read:2002hq} is adopted using the \textsc{RooStats}
package \cite{RooStats}.  The chosen test statistic, used to determine
how signal- and background-like the data are, is based on the profile
likelihood ratio. Systematic uncertainties are incorporated in the
analysis via nuisance parameters and treated as pseudo-observables,
following the frequentist paradigm. These uncertainties have been
listed in Section~\ref{sec:Systematics}.

The observed and the median expected 95\% confidence level (CL) limits as a function of the Higgs boson mass are shown in
Fig.~\ref{fig:xsec_mA_bands_csvt} and listed in Table~\ref{tab:limits:xsec} in~\ref{sec:app:limits}. The 1$\sigma$ and
2$\sigma$ bands of the test statistic, including systematic
uncertainties, are also shown.

\subsection{Interpretation within the MSSM}
The cross section limits shown in  Fig.~\ref{fig:xsec_mA_bands_csvt} are
further translated into exclusion limits on the MSSM parameters
$\tan\beta$ and \mA.  The cross sections obtained with the
four-flavor NLO QCD calculation~\cite{Dittmaier:2003ej, Dawson:2003kb} and the 
five-flavor NNLO QCD calculation as implemented in
\textsc{bbh@nnlo}~\cite{Harlander:2003ai} for
$\PQb +  \mathrm{h/H/A}$ associated production have been combined using the Santander matching 
scheme~\cite{Harlander:2011aa}. The
branching fractions were computed with the
 \textsc{FeynHiggs}~\cite{Degrassi:2002fi,Frank:2006yh,Heinemeyer:1998yj,Heinemeyer:1998np}
and \textsc{HDECAY}~\cite{Djouadi:1997yw,Djouadi:2006bz} programs as described in Ref.~\cite{Heinemeyer:2013tqa}.

\begin{figure}[htbp]
  \centering
  \includegraphics[width=0.48\textwidth]{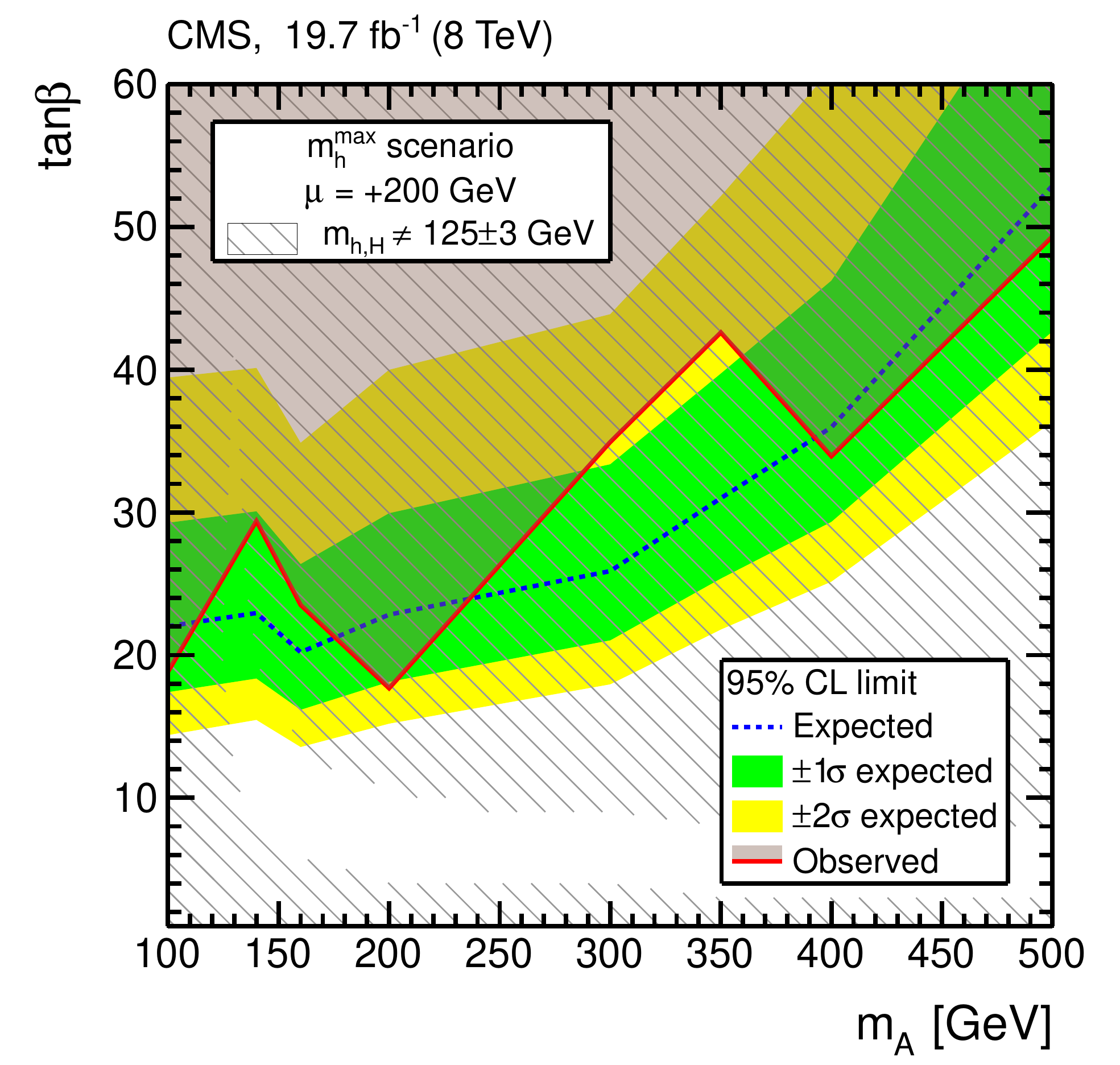}\hfill
  \includegraphics[width=0.48\textwidth]{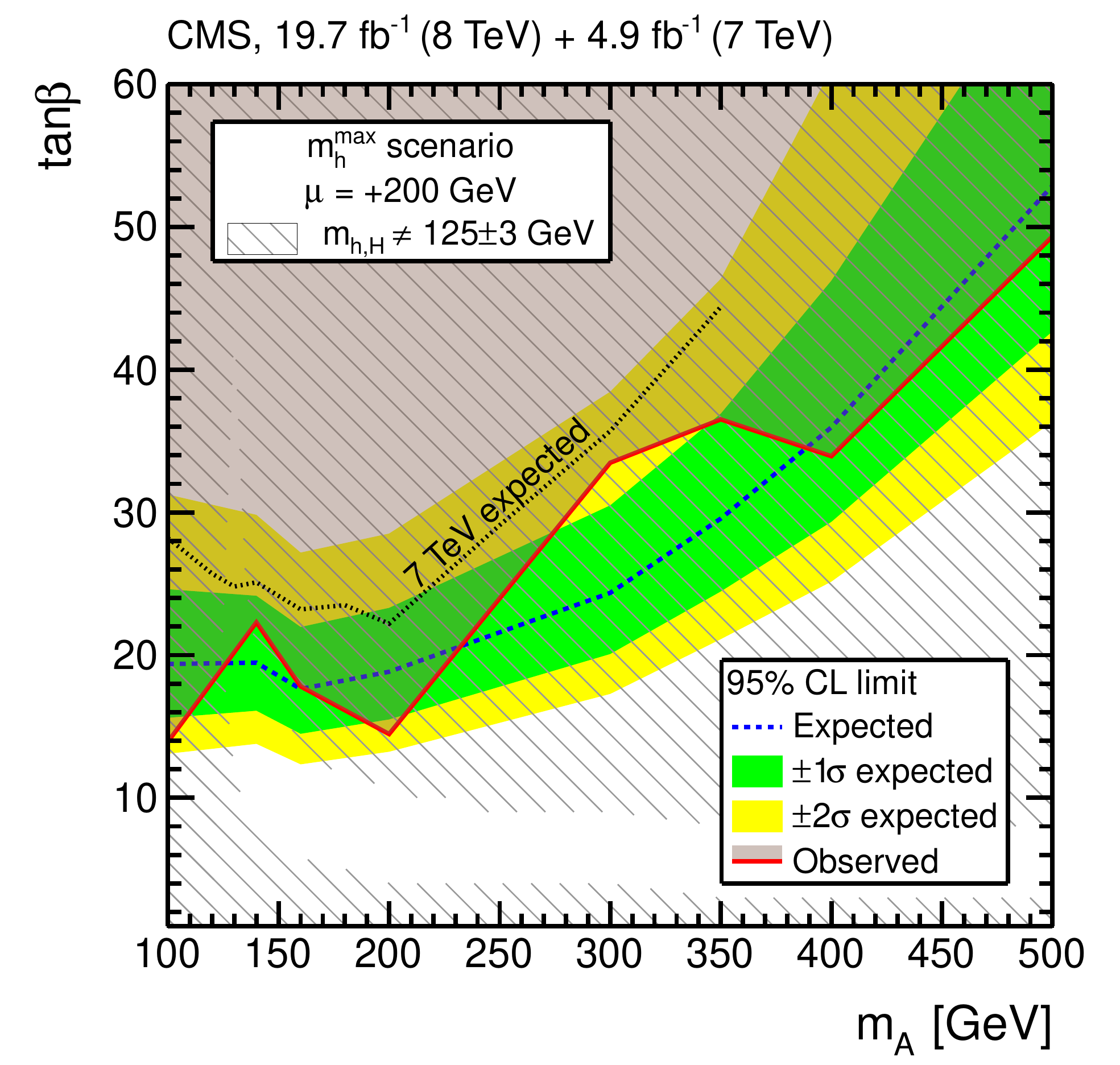}\hfill
  \caption{Expected and observed upper limits at \clsnf\ for
    the MSSM parameter $\tan \beta$ versus \mA in the \mhmax benchmark scenario with $\mu=+200\GeV$.
    The excluded parameter space (observed limit) is indicated by the shaded area.
    Regions where the mass of neither of the CP-even MSSM Higgs bosons h or H is compatible with the discovered Higgs boson of 125\GeV within a range of 3\GeV are marked by the hatched areas.
    The left plot shows the result obtained with the 8\TeV data only, the right plot shows the result obtained after a combination with the 7\TeV data.
    For comparison, the expected limit of the 7\TeV data analysis~\cite{Chatrchyan:2013qga} is overlayed.
  }
  \label{fig:tanb_mA_mhmax}
\end{figure}

The observed and expected \clsnf\ median upper limits on $\tan\beta$
versus \mA, together with the 1$\sigma$ and 2$\sigma$ bands, are
shown in Fig.~\ref{fig:tanb_mA_mhmax}~(left). They have been
computed within the traditional
MSSM~\mhmax benchmark scenario~\cite{Carena:2005ek} with the higgsino mass parameter $\mu = +200\GeV$.
The observed
upper limits range from $\tan\beta$ about 20 in the low-\mA region to about 50 at
$\mA=500\GeV$, and extend the existing measurement at
7\TeV~\cite{Chatrchyan:2013qga} into the hitherto unexplored \mA
region beyond 350\GeV. The model interpretation is not
extended to higher masses above 500\GeV because the theoretical predictions
are not reliable for $\tan \beta$ much higher than 60.

While the cross section limits obtained from the 2011 and 2012 data cannot
be combined directly due to the different center-of-mass energies, such a
combination is possible for the model-dependent interpretation. The
resulting upper limits on $\tan\beta$ versus \mA from both data
periods are shown in Figure~\ref{fig:tanb_mA_mhmax}~(right). While the
sensitivity is significantly enhanced compared to the 7\TeV
analysis~\cite{Chatrchyan:2013qga} already up to 350\GeV, the addition
of the 7\TeV result visibly improves the sensitivity in the low-mass
area below 200\GeV. The observed limit for $\tan\beta$ ranges down to
about 14 at the lowest \mA value considered.

\begin{figure}[htbp]
  \centering
  \includegraphics[width=0.48\textwidth]{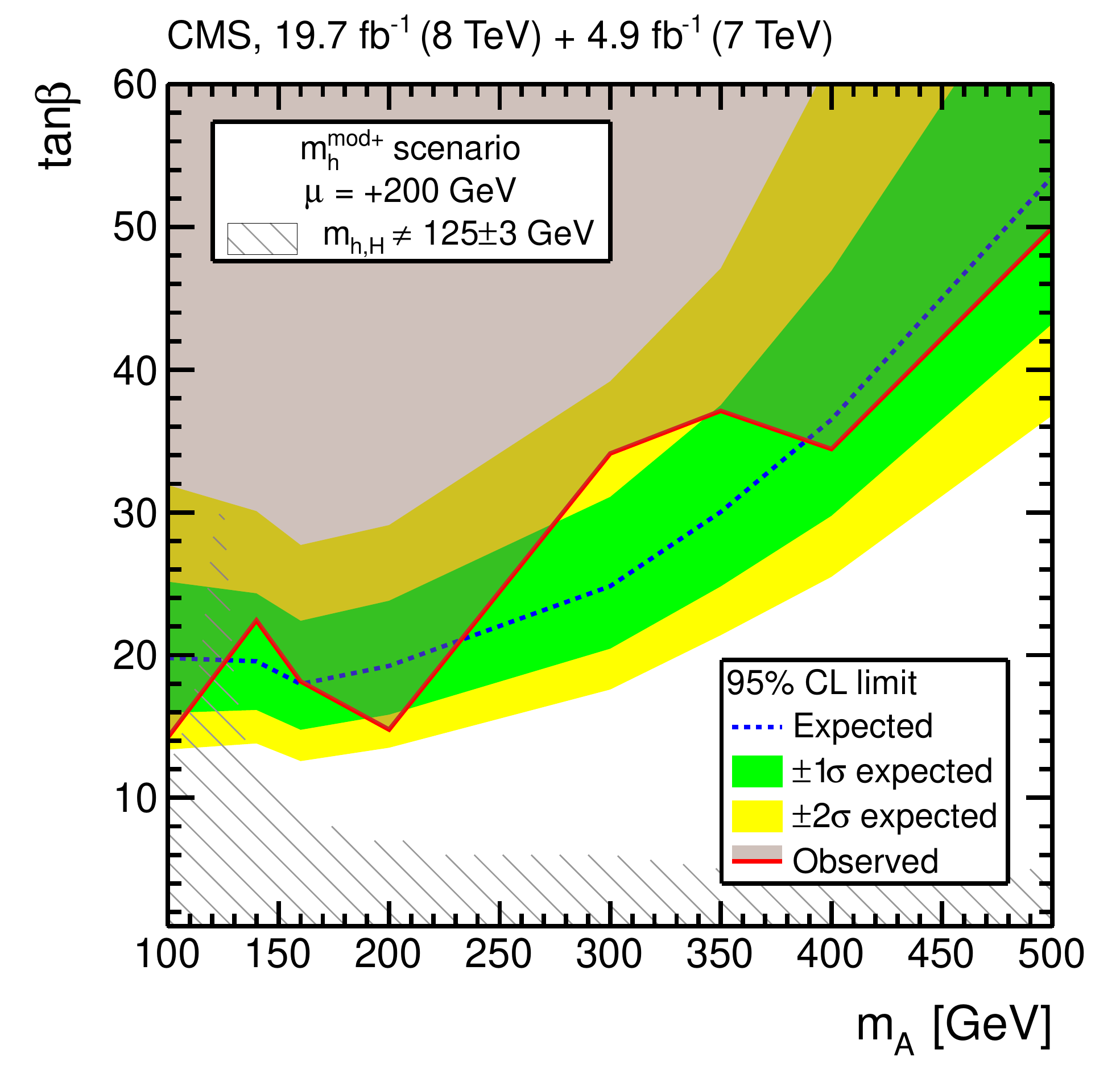}\hfill
  \includegraphics[width=0.48\textwidth]{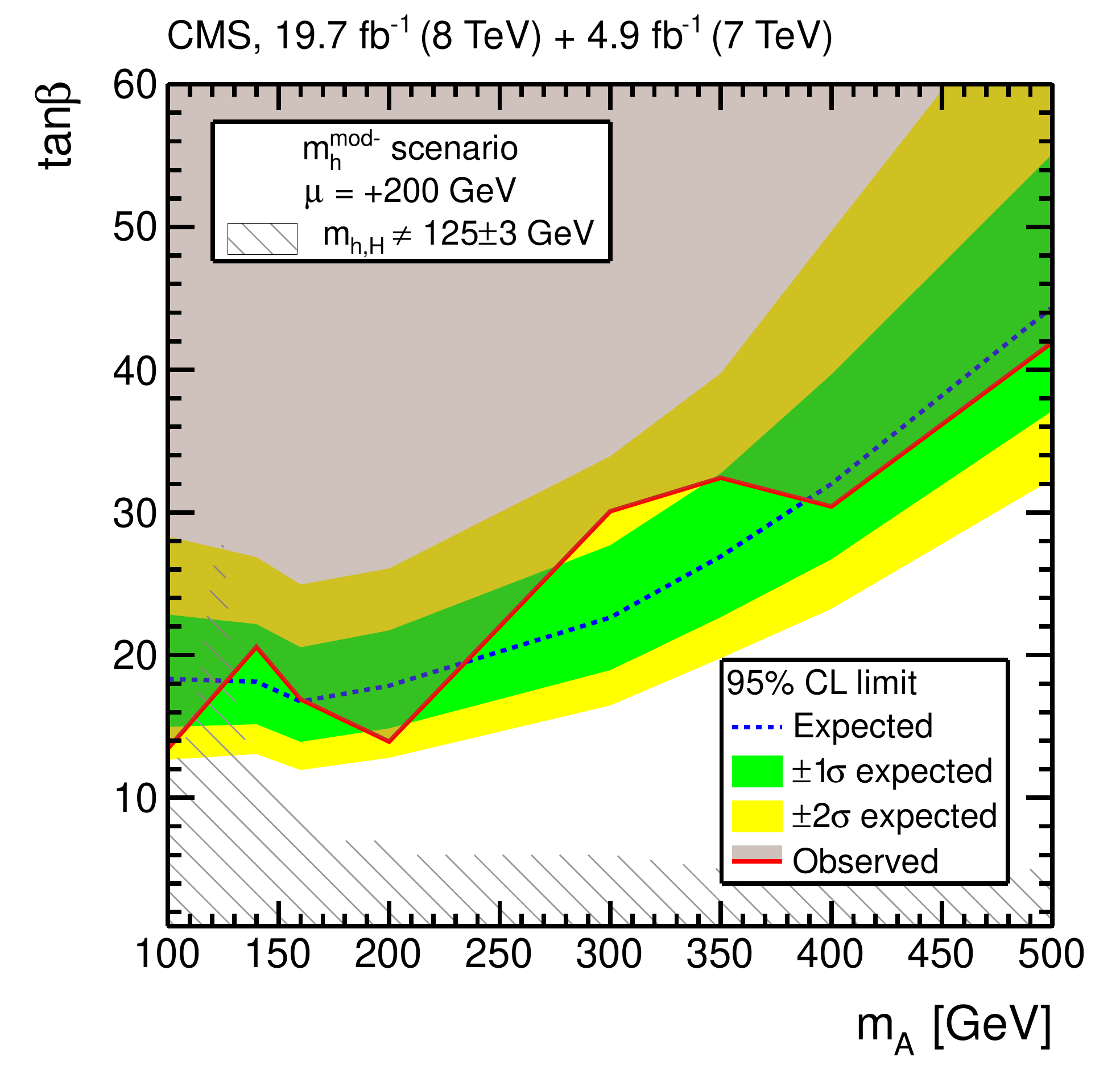}\hfill
  \includegraphics[width=0.48\textwidth]{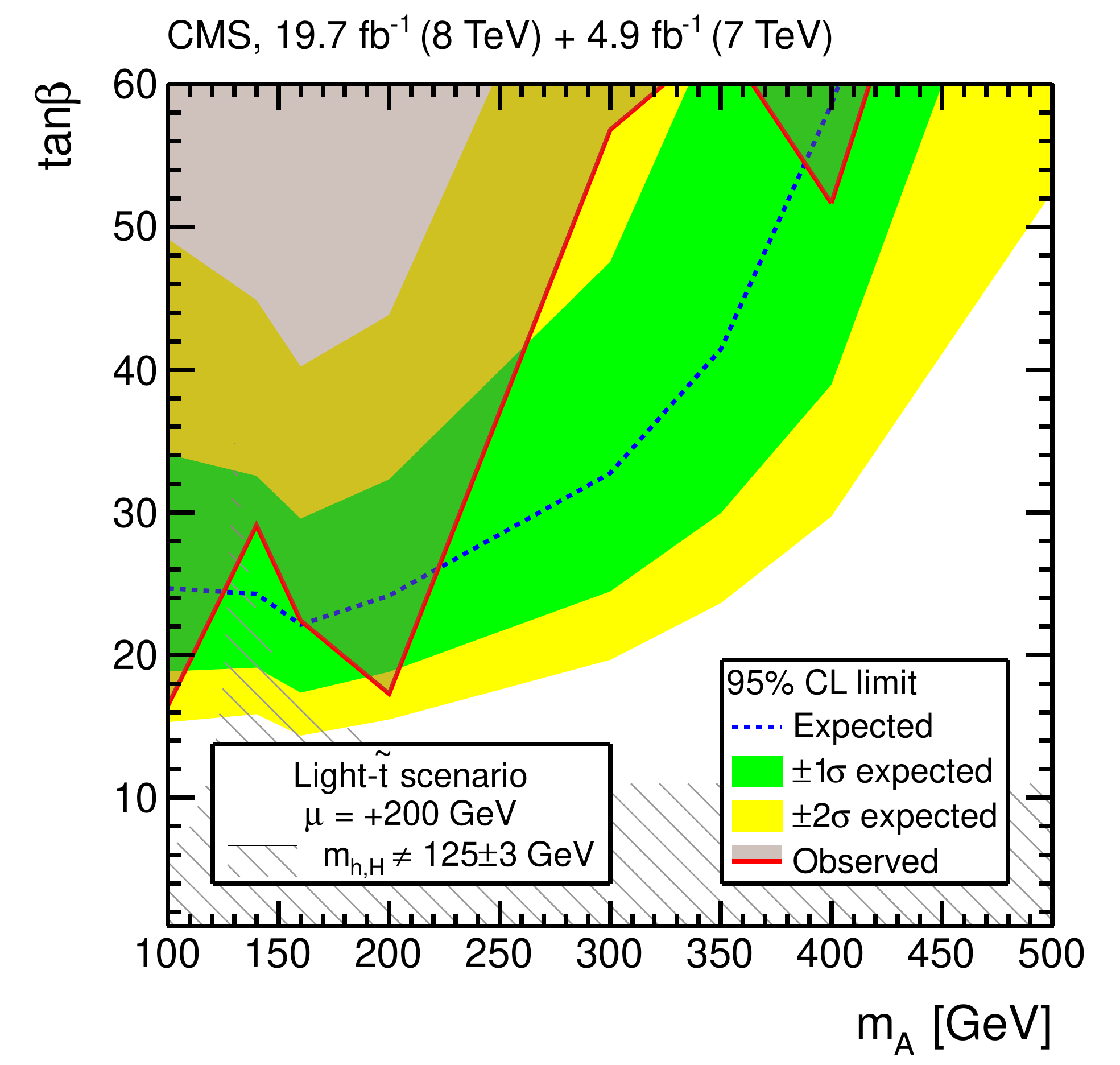}\hfill
  \includegraphics[width=0.48\textwidth]{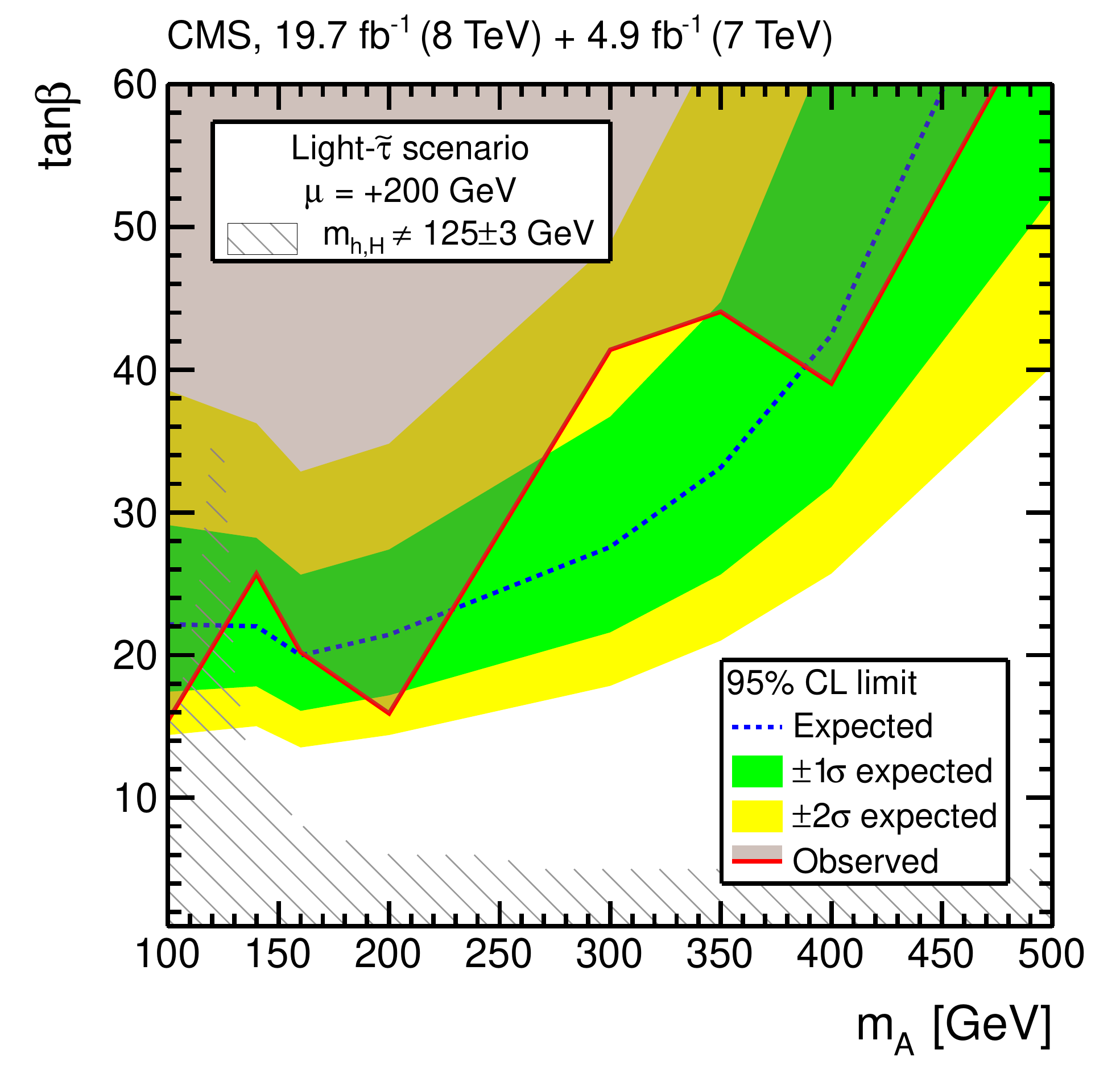}\hfill
  \caption{Expected and observed upper limits at \clsnf\ for
    the MSSM parameter $\tan\beta$ versus \mA in the \mhmodp, \mhmodm,
    light-stop, and light-stau benchmark scenarios with $\mu=+200$\GeV~\cite{Carena:2013qia}.
  }
  \label{fig:tanb_mA_newbs}
\end{figure}

Association of one of the CP-even MSSM Higgs bosons h and H with the measured state
at a mass of 125\GeV within a margin of $\pm 3\GeV$ that reflects the theoretical
uncertainties~\cite{Carena:2013qia} leads to an indirect constraint on $\tan\beta$.
The incompatible regions in the parameter space are illustrated by the hatched areas in both plots in
Fig.~\ref{fig:tanb_mA_mhmax}. In the \mhmax scenario, the MSSM
parameters beyond tree level have been tuned such that \mh
becomes as large as possible. As a result, large \mA and already
moderate values of $\tan\beta$ lead to \mh values that are
higher than the measured Higgs boson mass. This apparent exclusion of
large $\tan\beta$ values is, however, an artificial consequence of the
assumptions in the \mhmax scenario. Recently, several new MSSM
benchmark scenarios have been proposed, which are more naturally compatible with the
observed Higgs boson at
125\GeV~\cite{Carena:2013qia}, and among them the \mhmodp,
\mhmodm, light-stop, and light-stau scenarios are also used in the
following for the interpretation of the results of this analysis. The
observed and expected \clsnf\ exclusion limits in these scenarios
with $\mu=+200$\GeV, obtained with the
combined 7 and 8\TeV data, are shown in Fig.~\ref{fig:tanb_mA_newbs}.
(The term ``stop'' refers to the supersymmetric partner of the top quark throughout this paper.
Results for the $\tau$-phobic and low-\mH scenarios are not shown
because the analysis has sensitivity in a limited mass region only.)
The limits obtained in all MSSM benchmark scenarios are listed in
Tables~\ref{tab:limits:mhmax8TeV} to~\ref{tab:limits:lightstop} in
\ref{sec:app:limits}.

\begin{figure}[htbp]
  \centering
  \includegraphics[width=0.48\textwidth]{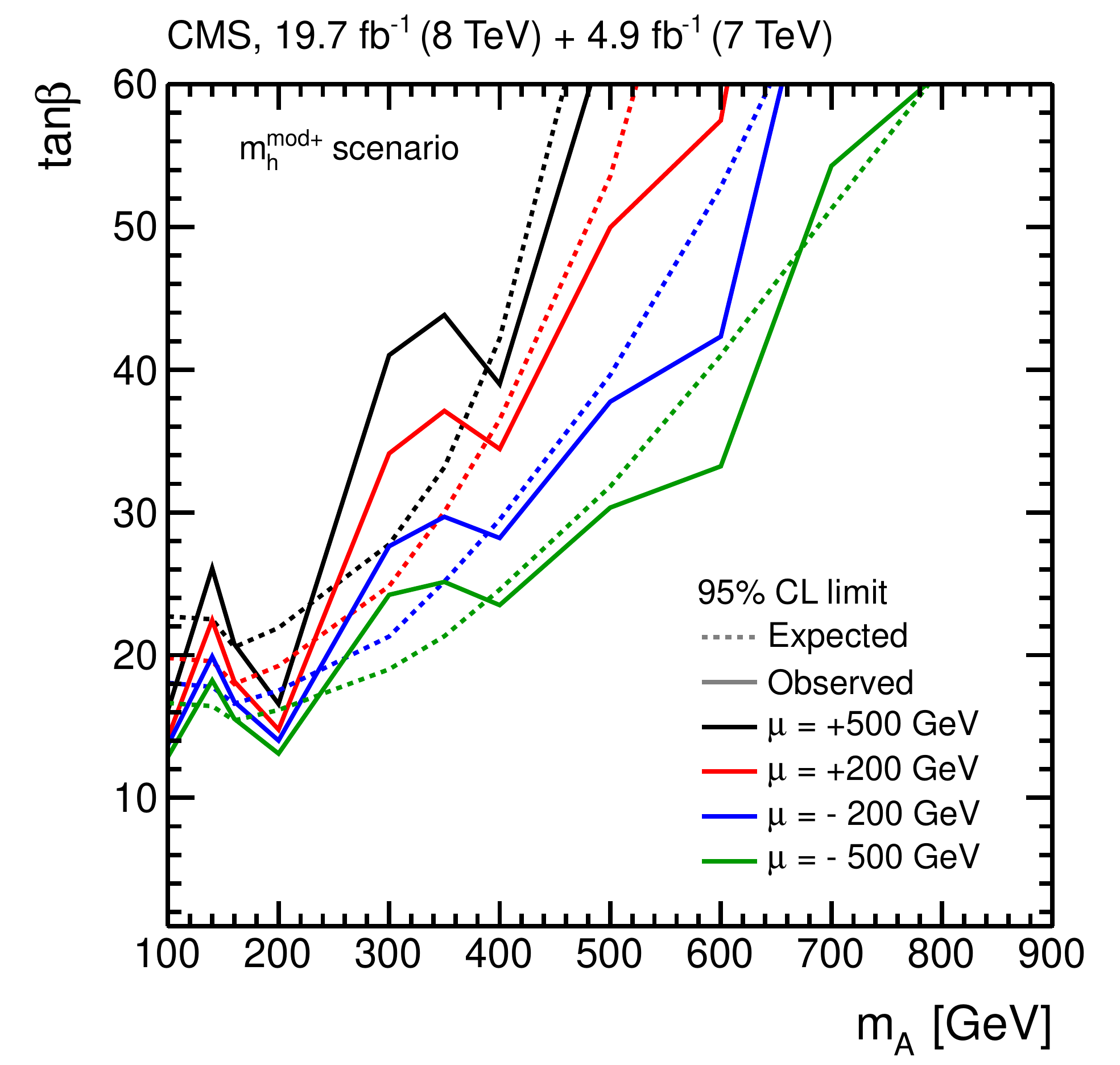}\hfill
  \includegraphics[width=0.48\textwidth]{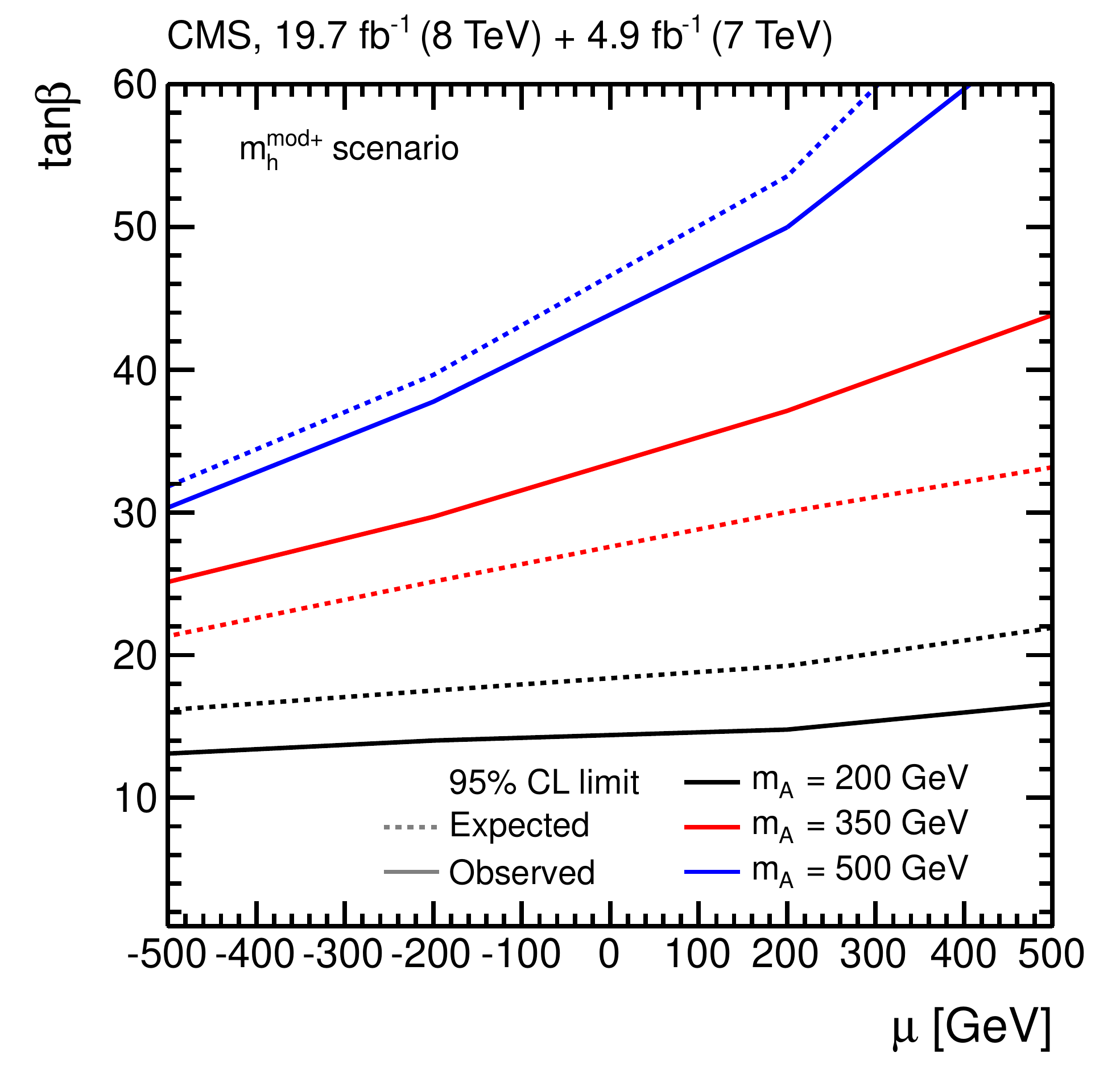}\hfill
  \caption{Expected and observed upper limits at \clsnf\ for the MSSM parameter $\tan\beta$ versus \mA for four different values of the higgsino mass parameter $\mu$ (left) and versus $\mu$ for three different values of \mA (right) in the \mhmodp scenario.
  }
  \label{fig:tanb_mA_vsmu}
\end{figure}

The aforementioned sensitivity of the $\phi\to\bbbar$ channel to the higgsino mass parameter $\mu$ is
evident in Fig.~\ref{fig:tanb_mA_vsmu}, where the limit in the \mhmodp scenario is compared for different
values of $\mu$. The dependence is particularly pronounced at higher \mA; for example, the observed upper
limit on $\tan\beta$ varies from 30 for $\mu=-500\GeV$ to beyond 60 for $\mu=+500\GeV$ for $\mA=500\GeV$.
The limits are also listed in Table~\ref{tab:limits:mhmodpvsmu} in \ref{sec:app:limits}.

\section{Summary}
A search for a Higgs boson decaying into a pair of b quarks and
accompanied by at least one additional b quark has been performed
in proton-proton collisions at a center-of-mass energy
of 8\TeV at the LHC, corresponding to an integrated luminosity
of 19.7\fbinv.
 The data were taken with dedicated triggers using
all-hadronic jet signatures combined with online b tagging. A
selection of events with three b-tagged jets has been performed in the
offline analysis. A signal has been searched for in the two-dimensional
spectrum formed by the invariant mass of the two leading jets and a condensed
event b-tag estimator.

No evidence for a signal is found. The observed distributions are well described
by a background model constructed from events in which only two of the
three leading jets are required to be b tagged. Upper limits on the Higgs boson cross
section times branching fraction are obtained in the mass region from
100--900\GeV, thus extending the search to considerably higher masses than those accessed
by the previous 7\TeV analysis.  The upper limits range from
about 250\unit{pb} at the lower end of the mass range, to about 1\unit{pb} at
900\GeV.

The results are interpreted within the MSSM in the benchmark
scenarios $\mhmax$, $\mhmodp$, $\mhmodm$,
light-stau and light-stop, and lead to upper limits for the model parameter $\tan\beta$
as a function of the mass parameter \mA. In combination with the 7\TeV
data, the observed limit for $\tan\beta$ ranges down to about 14
at the lowest \mA value of 100\GeV in the $\mhmodp$ scenario with a higgsino mass parameter of $\mu=+200\GeV$.
The limit depends significantly on $\mu$, varying from $\tan\beta=30$ for $\mu=-500\GeV$ to beyond 60 for $\mu=+500\GeV$ at $\mA=500\GeV$.

\begin{acknowledgments}
We congratulate our colleagues in the CERN accelerator departments for the excellent performance of the LHC and thank the technical and administrative staffs at CERN and at other CMS institutes for their contributions to the success of the CMS effort. In addition, we gratefully acknowledge the computing centers and personnel of the Worldwide LHC Computing Grid for delivering so effectively the computing infrastructure essential to our analyses. Finally, we acknowledge the enduring support for the construction and operation of the LHC and the CMS detector provided by the following funding agencies: BMWFW and FWF (Austria); FNRS and FWO (Belgium); CNPq, CAPES, FAPERJ, and FAPESP (Brazil); MES (Bulgaria); CERN; CAS, MoST, and NSFC (China); COLCIENCIAS (Colombia); MSES and CSF (Croatia); RPF (Cyprus); MoER, ERC IUT and ERDF (Estonia); Academy of Finland, MEC, and HIP (Finland); CEA and CNRS/IN2P3 (France); BMBF, DFG, and HGF (Germany); GSRT (Greece); OTKA and NIH (Hungary); DAE and DST (India); IPM (Iran); SFI (Ireland); INFN (Italy); MSIP and NRF (Republic of Korea); LAS (Lithuania); MOE and UM (Malaysia); CINVESTAV, CONACYT, SEP, and UASLP-FAI (Mexico); MBIE (New Zealand); PAEC (Pakistan); MSHE and NSC (Poland); FCT (Portugal); JINR (Dubna); MON, RosAtom, RAS and RFBR (Russia); MESTD (Serbia); SEIDI and CPAN (Spain); Swiss Funding Agencies (Switzerland); MST (Taipei); ThEPCenter, IPST, STAR and NSTDA (Thailand); TUBITAK and TAEK (Turkey); NASU and SFFR (Ukraine); STFC (United Kingdom); DOE and NSF (USA).

Individuals have received support from the Marie-Curie program and the European Research Council and EPLANET (European Union); the Leventis Foundation; the A. P. Sloan Foundation; the Alexander von Humboldt Foundation; the Belgian Federal Science Policy Office; the Fonds pour la Formation \`a la Recherche dans l'Industrie et dans l'Agriculture (FRIA-Belgium); the Agentschap voor Innovatie door Wetenschap en Technologie (IWT-Belgium); the Ministry of Education, Youth and Sports (MEYS) of the Czech Republic; the Council of Science and Industrial Research, India; the HOMING PLUS program of the Foundation for Polish Science, cofinanced from European Union, Regional Development Fund; the Compagnia di San Paolo (Torino); the Consorzio per la Fisica (Trieste); MIUR project 20108T4XTM (Italy); the Thalis and Aristeia programs cofinanced by EU-ESF and the Greek NSRF; the National Priorities Research Program by Qatar National Research Fund; and Rachadapisek Sompot Fund for Postdoctoral Fellowship, Chulalongkorn University (Thailand).
\end{acknowledgments}

\bibliography{auto_generated}

\appendix
\renewcommand\thesection{Appendix \Alph{section}}
\section{Signal Efficiency} \label{sec:app:efficiency}
\begin{figure}[tp]
  \centering
    \includegraphics[width=0.8\textwidth]{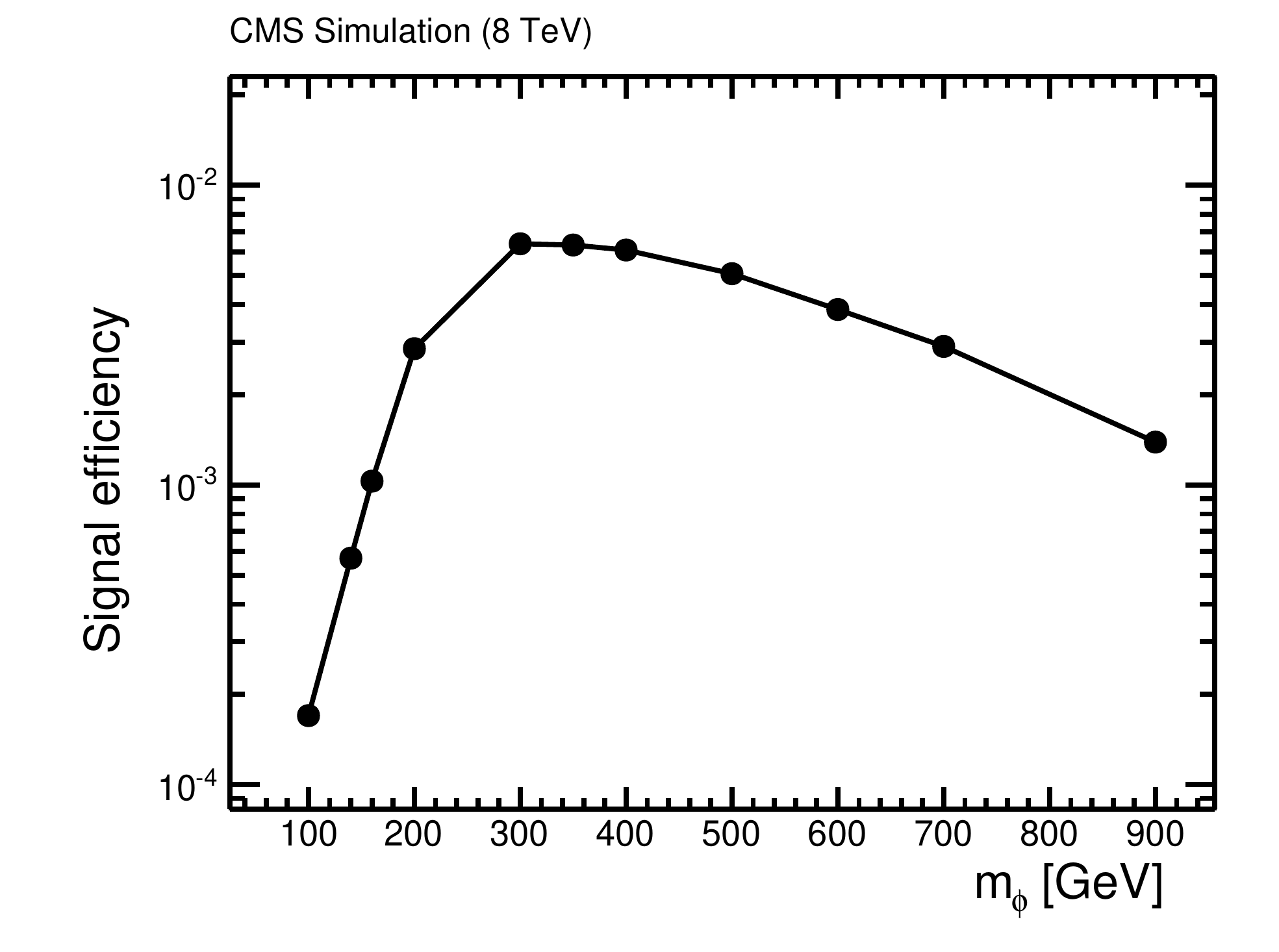}
\caption{\label{fig:signalEfficiency}
    The signal efficiency as a function of the Higgs boson mass $m_\phi$, for a center-of-mass energy of 8\TeV.
  }
\end{figure}

The signal efficiencies are
 summarized in Table~\ref{tab:signalEfficiency} and
shown in Fig.~\ref{fig:signalEfficiency} as a function of the
Higgs boson mass.

\begin{table}[htbp]
  \topcaption{The total signal efficiency in per mille as a function of the Higgs boson mass
    \mPhi, for a center-of-mass energy of 8\TeV.}
  \label{tab:signalEfficiency}
  \centering
    \begin{tabular}{cc}
      \hline
      $m_{\phi}\;[\GeVns{}]$ & Efficiency [per mille] \\
      \hline
      100                & 0.17 \\
      140                & 0.57 \\
      160                & 1.03 \\
      200                & 2.85 \\
      300                & 6.38 \\
      350                & 6.32 \\
      400                & 6.08 \\
      500                & 5.07 \\
      600                & 3.85 \\
      700                & 2.90 \\
      900                & 1.39 \\
      \hline
    \end{tabular}

\end{table}

\section{Exclusion limits} \label{sec:app:limits}

The model-independent \clsnf\ limits on $\sigma(\Pp\Pp \to \PQb\phi + \mathrm{X})
\, \mathcal{B}(\phi \to \bbbar)$ are listed in
Table~\ref{tab:limits:xsec} for different Higgs boson masses \mPhi.
The \clsnf\ limits of $(\tan\beta,\mA)$ are listed in Tables~\ref{tab:limits:mhmax8TeV}
to~\ref{tab:limits:lightstop} for different MSSM benchmark scenarios with $\mu=+200\GeV$
and for different values of $\mu$ in the \mhmodp scenario in Table~\ref{tab:limits:mhmodpvsmu}.

\begin{table}[htb]
  \topcaption{Expected and observed \clsnf\ upper limits on $\sigma(\Pp\Pp\to\PQb\phi+\mathrm{X})
\,\mathcal{B}(\phi\to\bbbar)$ in pb as a function of \mPhi,
where $\phi$ denotes a generic Higgs-like state, as obtained
from the 8\TeV data.}
  \label{tab:limits:xsec}
  \centering
    \begin{tabular}{crrrrrr}
      \hline
      Mass [\GeVns{}] &   $-2\sigma$ &   $-1\sigma$ & Median &   $+1\sigma$ &   $+2\sigma$ & Observed \\
      \hline
      100      & 160.7 & 221.0 &  330.4 & 518.8 & 811.2 &    251.9 \\
      140      &  49.4 &  68.0 &  101.6 & 161.2 & 254.7 &    158.8 \\
      160      &  25.9 &  35.6 &   52.5 &  81.6 & 126.1 &     68.7 \\
      200      &  13.7 &  19.0 &   28.2 &  44.1 &  68.6 &     17.8 \\
      300      &   3.0 &   4.1 &    6.1 &   9.4 &  14.5 &     10.5 \\
      350      &   1.9 &   2.7 &    3.9 &   6.1 &   9.3 &      7.1 \\
      400      &   1.3 &   1.8 &    2.7 &   4.2 &   6.4 &      2.4 \\
      500      &   0.8 &   1.2 &    1.7 &   2.7 &   4.2 &      1.5 \\
      600      &   0.6 &   0.9 &    1.3 &   2.0 &   3.2 &      0.7 \\
      700      &   0.5 &   0.7 &    1.1 &   1.8 &   2.8 &      1.3 \\
      900      &   0.4 &   0.6 &    1.0 &   1.6 &   2.7 &      0.8 \\
      \hline
    \end{tabular}

\end{table}

\begin{table}[htb]
  \topcaption{Expected and observed \clsnf\ upper limits on $\tan\beta$ as a function of \mA in the \mhmax,
$\mu = +200\GeV$, benchmark scenario obtained from the 8\TeV data only.}
  \label{tab:limits:mhmax8TeV}
  \centering
    \begin{tabular}{crrrrrr}
      \hline
      Mass [\GeVns{}] &  $-2\sigma$ &  $-1\sigma$ & Median &  $+1\sigma$ &  $+2\sigma$ & Observed \\
      \hline
      100      & 14.4 & 17.4 &   22.0 & 29.3 & 39.4 &     18.8 \\
      140      & 15.5 & 18.4 &   22.9 & 30.1 & 40.1 &     29.4 \\
      160      & 13.6 & 16.2 &   20.2 & 26.4 & 34.9 &     23.5 \\
      200      & 15.2 & 18.1 &   22.8 & 29.9 & 40.0 &     17.7 \\
      300      & 18.0 & 21.0 &   25.9 & 33.4 & 43.9 &     34.9 \\
      350      & 21.8 & 25.4 &   31.0 & 39.7 & 52.2 &     42.6 \\
      400      & 25.1 & 29.3 &   36.0 & 46.2 &  --- &     33.9 \\
      500      & 36.4 & 42.7 &   52.9 &  --- &  --- &     49.4 \\
      \hline
    \end{tabular}

\end{table}

\begin{table}[htb]
  \topcaption{Expected and observed \clsnf\ upper limits on $\tan\beta$ as a function of \mA in the \mhmax,
$\mu = +200\GeV$, benchmark scenario obtained from a combination of the 7 and 8\TeV data.}
  \label{tab:limits:mhmax}
  \centering
    \begin{tabular}{crrrrrr}
      \hline
      Mass [\GeVns{}] &  $-2\sigma$ &  $-1\sigma$ & Median &  $+1\sigma$ &  $+2\sigma$ & Observed \\
      \hline
      100      & 13.1 & 15.6 &   19.4 & 24.6 & 31.2 &     13.9 \\
      140      & 13.8 & 16.1 &   19.5 & 24.2 & 29.8 &     22.3 \\
      160      & 12.3 & 14.5 &   17.6 & 22.0 & 27.2 &     17.8 \\
      200      & 13.2 & 15.5 &   18.8 & 23.3 & 28.5 &     14.5 \\
      300      & 17.3 & 20.1 &   24.4 & 30.5 & 38.5 &     33.5 \\
      350      & 21.1 & 24.5 &   29.6 & 36.9 & 46.4 &     36.5 \\
      400      & 25.1 & 29.3 &   36.0 & 46.2 &  --- &     33.9 \\
      500      & 36.4 & 42.7 &   52.9 &  --- &  --- &     49.4 \\
      \hline
    \end{tabular}

\end{table}

\begin{table}[htb]
  \topcaption{Expected and observed \clsnf\ upper limits on $\tan\beta$ as a function of \mA in the \mhmodp,
$\mu = +200\GeV$, benchmark scenario obtained from a combination of the 7 and 8\TeV data.}
  \label{tab:limits:mhmodp}
  \centering
    \begin{tabular}{crrrrrr}
      \hline
      Mass [\GeVns{}] &  $-2\sigma$ &  $-1\sigma$ & Median &  $+1\sigma$ &  $+2\sigma$ & Observed \\
      \hline
      100      & 13.4 & 16.0 &   19.8 & 25.1 & 31.9 &     14.2 \\
      140      & 13.8 & 16.2 &   19.6 & 24.3 & 30.1 &     22.4 \\
      160      & 12.6 & 14.8 &   18.0 & 22.4 & 27.7 &     18.2 \\
      200      & 13.5 & 15.8 &   19.2 & 23.8 & 29.1 &     14.8 \\
      300      & 17.6 & 20.5 &   24.8 & 31.1 & 39.2 &     34.1 \\
      350      & 21.4 & 24.8 &   30.0 & 37.5 & 47.1 &     37.1 \\
      400      & 25.5 & 29.8 &   36.5 & 46.9 &  --- &     34.4 \\
      500      & 36.8 & 43.2 &   53.5 &  --- &  --- &     50.0 \\
      \hline
    \end{tabular}

\end{table}

\begin{table}[htb]
  \topcaption{Expected and observed \clsnf\ upper limits on $\tan\beta$ as a function of \mA in the \mhmodm,
$\mu = +200\GeV$, benchmark scenario obtained from a combination of the 7 and 8\TeV data.}
  \label{tab:limits:mhmodm}
  \centering
    \begin{tabular}{crrrrrr}
      \hline
      Mass [\GeVns{}] &  $-2\sigma$ &  $-1\sigma$ & Median &  $+1\sigma$ &  $+2\sigma$ & Observed \\
      \hline
      100      & 12.7 & 15.0 &   18.3 & 22.8 & 28.3 &     13.4 \\
      140      & 13.1 & 15.2 &   18.1 & 22.2 & 26.9 &     20.6 \\
      160      & 11.9 & 13.9 &   16.7 & 20.5 & 24.9 &     16.9 \\
      200      & 12.8 & 14.9 &   17.9 & 21.7 & 26.1 &     13.9 \\
      300      & 16.5 & 18.9 &   22.6 & 27.7 & 33.9 &     30.1 \\
      350      & 19.8 & 22.7 &   26.9 & 32.7 & 39.7 &     32.4 \\
      400      & 23.2 & 26.7 &   32.0 & 39.7 & 49.7 &     30.4 \\
      500      & 32.3 & 37.1 &   44.4 & 55.1 &  --- &     41.9 \\
      \hline
    \end{tabular}

\end{table}

\begin{table}[htb]
  \topcaption{Expected and observed \clsnf\ upper limits on $\tan\beta$ as a function of \mA in the light-stau, $\mu = +200\GeV$, benchmark scenario obtained from a combination of the 7 and 8\TeV data.}
  \label{tab:limits:lighstau}
  \centering
    \begin{tabular}{crrrrrr}
      \hline
      Mass [\GeVns{}] &  $-2\sigma$ &  $-1\sigma$ & Median &  $+1\sigma$ &  $+2\sigma$ & Observed \\
      \hline
      100      & 14.4 & 17.4 &   22.2 & 29.1 & 38.6 &     15.4 \\
      140      & 15.0 & 17.8 &   22.0 & 28.2 & 36.2 &     25.7 \\
      160      & 13.5 & 16.1 &   20.0 & 25.6 & 32.9 &     20.2 \\
      200      & 14.4 & 17.2 &   21.4 & 27.4 & 34.8 &     15.9 \\
      300      & 17.8 & 21.6 &   27.6 & 36.7 & 48.9 &     41.4 \\
      350      & 21.0 & 25.7 &   33.1 & 44.8 &  --- &     44.1 \\
      400      & 25.7 & 31.8 &   42.4 &  --- &  --- &     39.0 \\
      500      & 40.3 & 52.1 &    --- &  --- &  --- &      --- \\
      \hline
    \end{tabular}

\end{table}

\begin{table}[htb]
  \topcaption{Expected and observed \clsnf\ upper limits on $\tan\beta$ as a function of \mA in the light-stop, $\mu = +200\GeV$, benchmark scenario obtained from a combination of the 7 and 8\TeV data.}
  \label{tab:limits:lightstop}
  \centering
    \begin{tabular}{crrrrrr}
      \hline
      Mass [\GeVns{}] &  $-2\sigma$ &  $-1\sigma$ & Median &  $+1\sigma$ &  $+2\sigma$ & Observed \\
      \hline
      100      & 15.3 & 18.9 &   24.7 & 34.1 & 49.2 &     16.5 \\
      140      & 15.9 & 19.1 &   24.3 & 32.6 & 44.9 &     29.1 \\
      160      & 14.4 & 17.4 &   22.1 & 29.6 & 40.2 &     22.4 \\
      200      & 15.5 & 18.8 &   24.2 & 32.3 & 43.8 &     17.3 \\
      300      & 19.7 & 24.5 &   32.7 & 47.6 &  --- &     56.8 \\
      350      & 23.6 & 29.9 &   41.4 &  --- &  --- &      --- \\
      400      & 29.7 & 39.0 &   58.6 &  --- &  --- &     51.7 \\
      500      & 52.5 &  --- &    --- &  --- &  --- &      --- \\
      \hline
    \end{tabular}
  
\end{table}

\begin{table}[htb]
  \topcaption{Observed (expected) \clsnf\ upper limits on $\tan\beta$ as a function of \mA in the \mhmodp benchmark scenario for different values of the higgsino mass parameter $\mu$ obtained from a combination of the 7 and 8\TeV data.}
  \label{tab:limits:mhmodpvsmu}
  \centering
    \begin{tabular}{cr@{\;}lr@{\;}lr@{\;}lr@{\;}l}
      \hline
      Mass [\GeVns{}] & \multicolumn{2}{c}{$\mu=-500\GeV$} &  \multicolumn{2}{c}{$\mu=-200\GeV$} & \multicolumn{2}{c}{$\mu=+200\GeV$} & \multicolumn{2}{c}{$\mu=+500\GeV$} \\
      \hline
      100 & 12.9 & (16.6) & 13.7 & (18.1) & 14.2 & (19.8) & 16.1 & (22.7)  \\
      140 & 18.2 & (16.4) & 19.9 & (17.8) & 22.4 & (19.6) & 26.1 & (22.5)  \\
      160 & 15.5 & (15.4) & 16.7 & (16.6) & 18.2 & (18.0) & 20.8 & (20.6)  \\
      200 & 13.1 & (16.2) & 14.0 & (17.5) & 14.8 & (19.2) & 16.6 & (21.9)  \\
      300 & 24.2 & (19.0) & 27.6 & (21.3) & 34.1 & (24.8) & 41.0 & (27.8)  \\
      350 & 25.1 & (21.3) & 29.7 & (25.1) & 37.1 & (30.0) & 43.8 & (33.2)  \\
      400 & 23.5 & (24.6) & 28.2 & (29.5) & 34.4 & (36.5) & 39.0 & (42.2)  \\
      500 & 30.3 & (31.8) & 37.8 & (39.6) & 50.0 & (53.5) &  --- & (---)  \\
      600 & 33.2 & (41.0) & 42.3 & (52.8) & 57.5 &  (---) &  --- & (---)  \\
      700 & 54.3 & (51.3) &  --- &  (---) &  --- &  (---) &  --- & (---)  \\
      \hline
    \end{tabular}

\end{table}

\cleardoublepage \section{The CMS Collaboration \label{app:collab}}\begin{sloppypar}\hyphenpenalty=5000\widowpenalty=500\clubpenalty=5000\textbf{Yerevan Physics Institute,  Yerevan,  Armenia}\\*[0pt]
V.~Khachatryan, A.M.~Sirunyan, A.~Tumasyan
\vskip\cmsinstskip
\textbf{Institut f\"{u}r Hochenergiephysik der OeAW,  Wien,  Austria}\\*[0pt]
W.~Adam, E.~Asilar, T.~Bergauer, J.~Brandstetter, E.~Brondolin, M.~Dragicevic, J.~Er\"{o}, M.~Flechl, M.~Friedl, R.~Fr\"{u}hwirth\cmsAuthorMark{1}, V.M.~Ghete, C.~Hartl, N.~H\"{o}rmann, J.~Hrubec, M.~Jeitler\cmsAuthorMark{1}, V.~Kn\"{u}nz, A.~K\"{o}nig, M.~Krammer\cmsAuthorMark{1}, I.~Kr\"{a}tschmer, D.~Liko, T.~Matsushita, I.~Mikulec, D.~Rabady\cmsAuthorMark{2}, B.~Rahbaran, H.~Rohringer, J.~Schieck\cmsAuthorMark{1}, R.~Sch\"{o}fbeck, J.~Strauss, W.~Treberer-Treberspurg, W.~Waltenberger, C.-E.~Wulz\cmsAuthorMark{1}
% ...
\vskip\cmsinstskip
% ...
1:~~Also at Vienna University of Technology, Vienna, Austria\\
2:~~Also at CERN, European Organization for Nuclear Research, Geneva, Switzerland\\
% ...

\end{sloppypar}
\end{document}